\documentclass[sigconf,screen]{acmart}
\usepackage{graphicx}
\usepackage{tikz}
\usepackage{tabularx}
\usepackage{multirow}
\usepackage{multicol}
\usepackage[xcdraw,table]{xcolor}
\usetikzlibrary{trees}
\usepackage{subfig}
\usepackage{longtable}
\usepackage{booktabs}

\definecolor{color1}{HTML}{4190A1}
\definecolor{color2}{HTML}{1C4955}
\definecolor{color3}{HTML}{CA9379}
\definecolor{color4}{HTML}{7197D6}
\AtBeginDocument{%
  \providecommand\BibTeX{{%
    \normalfont B\kern-0.5em{\scshape i\kern-0.25em b}\kern-0.8em\TeX}}}

\copyrightyear{2026}
\acmYear{2026}
\setcopyright{cc}
\setcctype{by}
\acmConference[XX'26]{XX}{April 13--17, 2026}{Barcelona, Spain}
\acmBooktitle{XX (XX'26), April 13--17, 2026, Barcelona, Spain}
\acmPrice{}
\acmDOI{XX}
\acmISBN{XX}

\begin{document}

\title[A Scoping Review on Privacy Policy's Visualization]{A Scoping Review and Guidelines on Privacy Policy's Visualization from an HCI Perspective}

% \affiliation{%
\author{Shuning Zhang}
\orcid{0000-0002-4145-117X}
\email{zsn23@mails.tsinghua.edu.cn}
\affiliation{%
  \institution{Tsinghua University}
  \city{Beijing}
  \country{China}
}

\author{Eve He}
\orcid{0009-0004-7275-6074}
\email{eveanny.hx@gmail.com}
\affiliation{%
  \institution{Independent Researcher}
  \city{Madison, WI}
  \country{United States}
}

\author{Sixing Tao}
\orcid{0009-0004-1024-1032}
\email{sixint@cs.washington.edu}
\affiliation{%
  \institution{University of Washington}
  \city{Seattle, WA}
  \country{United States}
}

\author{Yuting Yang}
\orcid{0009-0007-1351-8377}
\email{yutingy@umich.edu}
\affiliation{%
  \institution{University of Michigan}
  \city{Ann Arbor}
  \state{Michigan}
  \country{USA}
}

\author{Ying Ma}
\orcid{0000-0001-5413-0132}
\email{yima3@student.unimelb.edu.au}
\affiliation{%
  \department{School of Computing and Information Systems}
  \institution{The University of Melbourne}
  \city{Melbourne}
  \country{Australia} 
}

\author{Ailei Wang}
\orcid{0009-0006-4447-7005}
\email{wal22@mails.tsinghua.edu.cn}
\affiliation{%
  \institution{Tsinghua University}
  \city{Beijing}
  \country{China}
}

\author{Xin Yi}
\orcid{0000-0001-8041-7962}
\authornote{Corresponding author.}
\affiliation{
    \institution{Tsinghua University}
    \city{Beijing}
    \country{China}
}
\email{yixin@tsinghua.edu.cn}

\author{Hewu Li}
\orcid{0000-0002-6331-6542}
\affiliation{
    \institution{Tsinghua University}
    \city{Beijing}
    \country{China}
}
\email{lihewu@cernet.edu.cn}

% \affiliation{%

% \affiliation{%

% \affiliation{%

% \affiliation{%

% \affiliation{%

% \affiliation{%

% \affiliation{%


\renewcommand{\shortauthors}{Zhang et al.}


\begin{abstract}

Privacy Policies are a cornerstone of informed consent, yet a persistent gap exists between their legal intent and practical efficacy. Despite decades of Human-Computer Interaction (HCI) research proposing various visualizations, user comprehension remains low, and designs rarely see widespread adoption. To understand this landscape and chart a path forward, we synthesized 65 top-tier papers using a framework adapted from the user-centered design lifecycle. Our analysis presented findings of the field's evolution across four dimensions: (1) the trade-off between information load and decision efficacy, which demonstrates a shift from augmenting disclosures to prioritizing information condensation and cognitive load management to counter the inefficacy of comprehensive texts, (2) the co-evolutionary dynamic of design and automation, revealing that complex design ambitions such as context-awareness drove the need for advanced NLP, while recent LLM breakthroughs are enabling the semantic interpretation required to realize those designs, (3) the tension between generality and specificity, highlighting the divergence between standardized, cross-platform solutions and the increasing necessity for specialized, context-aware interaction patterns in IoT and immersive environments, and (4) balancing stakeholder opinions, which shows that visualization efficacy is constrained by the complex interplay of regulatory mandates, developer capabilities and provider incentives. We conclude by outlining four critical challenges for future research.

\end{abstract}

\begin{CCSXML}
<ccs2012>
   <concept>
       <concept_id>10002978.10003029</concept_id>
       <concept_desc>Security and privacy~Human and societal aspects of security and privacy</concept_desc>
       <concept_significance>500</concept_significance>
       </concept>
   <concept>
       <concept_id>10003120.10003121</concept_id>
       <concept_desc>Human-centered computing~Human computer interaction (HCI)</concept_desc>
       <concept_significance>300</concept_significance>
       </concept>
   <concept>
       <concept_id>10002978.10003029.10011703</concept_id>
       <concept_desc>Security and privacy~Usability in security and privacy</concept_desc>
       <concept_significance>500</concept_significance>
       </concept>
 </ccs2012>
\end{CCSXML}

\ccsdesc[500]{Security and privacy~Human and societal aspects of security and privacy}
\ccsdesc[300]{Human-centered computing~Human computer interaction (HCI)}
\ccsdesc[500]{Security and privacy~Usability in security and privacy}

\keywords{Privacy Policy, Demonstration, Literature Review, Human-Computer Interaction}

\begin{teaserfigure}
    \centering
    \includegraphics[width=1\linewidth]{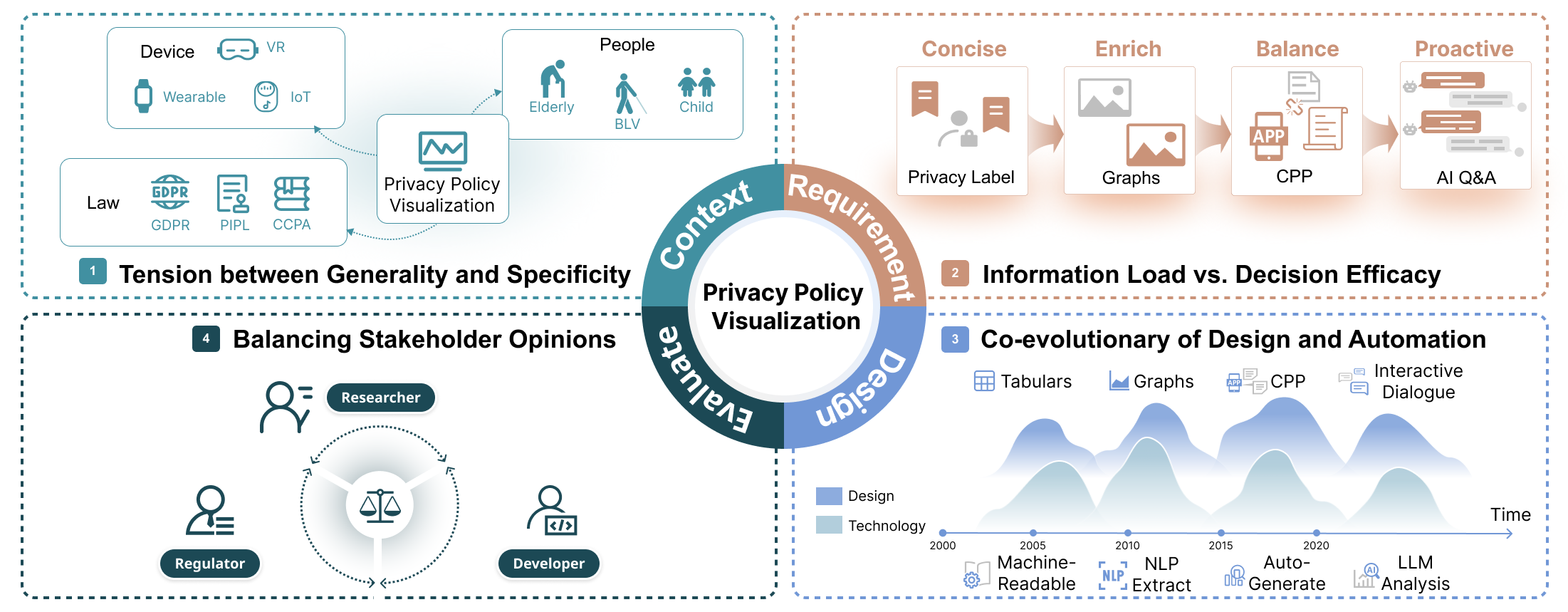}
    \caption{The framework of this paper, which consisted of four main findings, organized around four aspects: understanding \textit{context}, setting \textit{requirements}, \textit{designing} and \textit{evaluating}.}
    \Description{A conceptual framework diagram illustrating the lifecycle of privacy policy visualization research. The diagram is centered around a circular cycle labeled ``privacy policy visualization'' divided into four colored quadrants: Context, Requirement, design, and evaluate. The context quadrant, on the top-left connects icons of devices, such as wearables, IoT, and laws, such as GDPR, CCPA, to the central hub, labeled as ``tension between generality and specificity''. The requirement quadrant, on the top-right, shows a progression of interface icons labeled concise, enrich, balance and proactive, corresponding to a label ``information load versus decision efficacy''. The design quadrant, on the bottom-right, displays a timeline graph titled ``co-evolutionary of design and automation'', showing wave-like trends evolving from ``machine readable'' and ``tabulars'' to ``LLM analysis'' and ``interactive dialogue'' over time. The evaluate quadrant, on the bottom-left, depicts a triangular relationship between three stakeholder icons: ``regulator'', ``researcher'', and ``developer'', labeled ``balancing stakeholder opinions''.}
    \label{fig:teaser}
\end{teaserfigure}


\maketitle

\section{Introduction}

The communication of data handling practices is essential for digital trust and regulatory compliance, and it helps users to make informed decisions about their personal information~\cite{liccardi2014no,jensen2004privacy}. Privacy policies are the main mechanism for this purpose. They are intended to articulate an organization's data collection, usage, and protection protocols~\cite{jensen2004privacy}. However, a well-documented disconnect exists between the intended function of these policies and how effective they are in practice. Often presented as lengthy, opaque, and jargon-filled legal documents, they are often skipped and misunderstood by users, thus failing to achieve meaningful communication~\cite{jensen2004privacy,mcdonald2008cost,obar2020biggest}.

In response, Human-Computer Interaction (HCI) research has proposed various design visualizations, such as layering~\cite{mcdonald2009comparative}, graphical visualizations~\cite{reinhardt2021visual,tabassum2018increasing}, and just-in-time notices~\cite{windl2022automating,ortloff2020implementation}. Privacy policy visualization, containing both solely visualization (i.e., ``privacy notice'') and also controls (i.e., ``privacy control''), acts as a bridge between legal compliance and user agency across three dimensions: (1) \textbf{cognitive}, by translating opaque legal jargon into accessible mental models, (2) \textbf{behavioral}, by disrupting habitual policy avoidance through engaging interaction design, and (3) \textbf{regulatory}, by supporting the necessary conditions for valid, informed consent~\cite{liccardi2014no,jensen2004privacy,mcdonald2009comparative,reinhardt2021visual}. 

Crucially, the integration of Artificial Intelligence (AI), particularly Large Language Models (LLMs), propels new evolution for privacy policy visualization. Unlike traditional visualizations that rely on fixed summaries, AI agents offer the potential for generative interpretation~\cite{chen2025clear,freiberger2025you} and dynamic synthesis~\cite{zhang2025privcaptcha,freiberger2025you}. This capability opens new possibilities within the domain, suggesting potential transformations across visualization strategies~\cite{slate2025iterative}, automation paradigms~\cite{pan2024new}, and how different stakeholders balance their interests.~\cite{li2022understanding}.

However, despite this evolution, existing literature reviews have largely examined this domain through a static lens. Much prior work has analyzed the \textbf{policy text itself}, tracking its linguistic evolution, readability, and legal compliance~\cite{wagner2023privacy,amos2021privacy,belcheva2023understanding,javed2024systematic}. Another significant research stream has reviewed \textbf{computational approaches}, such as the use of Natural Language Processing (NLP) and LLMs for automated information extraction~\cite{del2022systematic}, or the development of formal machine-readable policy languages like Platform for Privacy Preferences (P3P)~\cite{kumaraguru2007survey,kasem2015security}. A third area synthesized the \textbf{user-centric problem}, confirming users' poor comprehension~\cite{raibulet2025awareness} and exploring the ``privacy paradox''~\cite{mitchell2022privacy,barth2021data}. While informative, these reviews overlooked the dynamic interplay among evolving challenges, technical capabilities, requirements, and design paradigms.

To address this gap, this paper adopts a temporal lens to examine the evolution of privacy policy visualization as a response to the shifting cognitive, technological and economic constraints. Here, we focus on works that both those work that visualize a policy, and those who also provide privacy controls, as static disclosures are evolving into interactive consent interfaces, increasingly blurring this distinction. \textbf{This evolutionary perspective is important for the HCI community, because it shows that many current design challenges are legacies of past limitations rather than isolated issues. By mapping this trajectory, we aim to uncover the rationale behind existing paradigms and provide a foundation for future design strategies.}

Therefore, we synthesize 65 papers from top-tier conferences to construct a systematic framework that charts the evolution of privacy policy visualization and distills actionable guidelines for its future development. Adopting a design lifecycle perspective~\cite{hasani2020user}, we examine the domain's trajectory, our analysis maps the field's evolution across four interconnected dimensions. We begin by analyzing (1) the productive tension between generality and specificity in \textbf{the context of use} (Section~\ref{sec:general_specific}), highlighting the divergence between generalized, cross-platform solutions and requirements for specialized, context-aware visualization in Internet of Things (IoT) and immersive environments. We then examine (2) the trade-offs between information load and design efficacy regarding \textbf{requirements} (Section~\ref{sec:load_efficiency}), showing a shift from condensing and balancing information, to fostering users' engagement and reflection. Subsequently, we analyze (3) the co-evolutionary relationship of \textbf{design solutions} and automation (Section~\ref{sec:design_automation}), revealing that design innovations drove the need for advanced NLP, while recent LLM breakthroughs enable new paradigms. Finally, we discuss (4) the multi-stakeholder dynamics involved in \textbf{evaluation} and deployment (Section~\ref{sec:multi_stakeholder}), which shows that deployment efficacy is constrained by the complex interplay among regulators, developers and end-users. The paper ends with a forward-looking summarization of four open challenges (Section~\ref{sec:discussion}).

\begin{figure}[h]
    \includegraphics[width=0.5\textwidth]{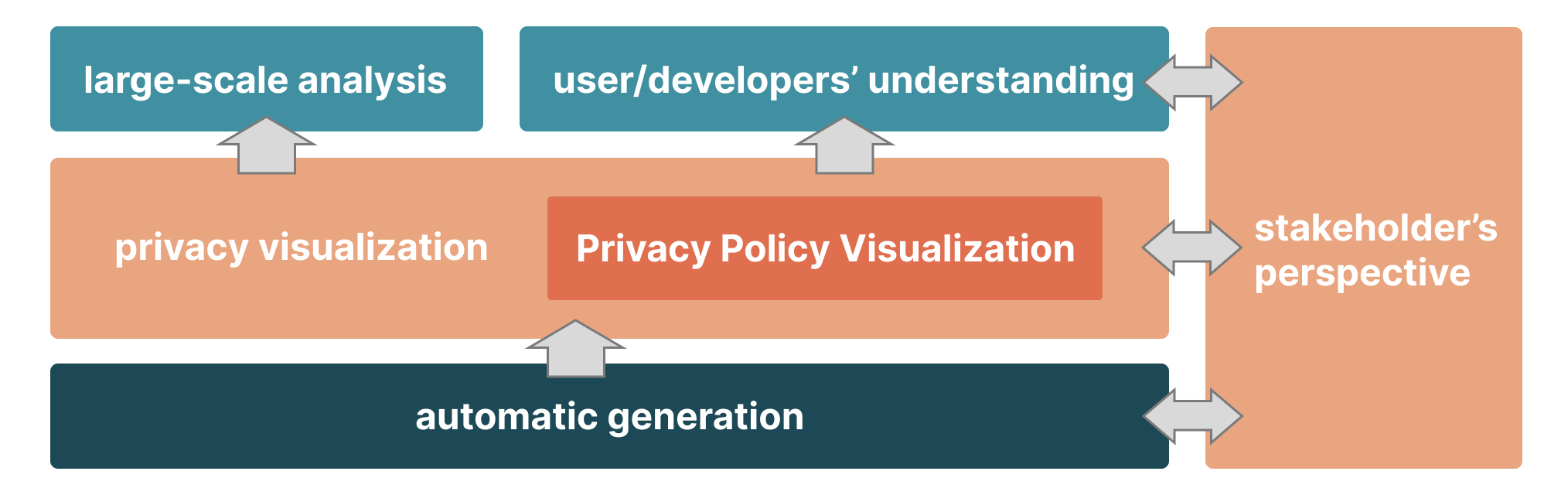}
    \caption{The scope of our paper, where privacy policy visualization is a sub-field of privacy visualization. Privacy policy visualization is closely interconnected to the understandings of users or developers, automatic generation, large-scale analysis and those work on stakeholders' perspective.}
    \label{fig:scope}
    \Description{A block diagram illustrating the scope of the literature review. A central, orange rectangular box labeled ``privacy visualization'' contains a nested, darker orange box labeled ``privacy policy visualization.'' This central block is connected via bi-directional arrows to three surrounding blocks: a blue block above labeled ``large-scale analysis'' and ``user/developers' understanding'', a dark blue block below labeled ``automatic generation'', and a vertical orange block to the right labeled ``stakeholder's perspective''. This visual arrangement indicates that privacy policy visualization is a sub-field influenced by and interacting with these adjacent research areas.}
    % The scope of our paper, where privacy policy visualization is a sub-field of privacy visualization. Privacy policy visualization is closely interconnected to the understandings of users or developers, automatic generation, large-scale analysis and those work on stakeholders' perspective.
\end{figure}

\section{Scope and Methodology}

This paper presents a scoping review of privacy policy visualization from an HCI perspective, focusing on the visualization's temporal evolution. We first delineate the review's scope (Section~\ref{sec:scope}), then motivate and present the research framework (Section~\ref{sec:framework}), and detail the systematic methodology for literature selection and analysis (Section~\ref{sec:methodology}). We finally summarize our contributions (Section~\ref{sec:contribution}). 

\subsection{Scope and Definition}\label{sec:scope}

\textbf{Privacy Visualization}: Researchers, regulators, and the industry call for visual representations of how online services handle personal data, so as to communicate privacy risks~\cite{barth2022understanding}. Privacy visualization could employ privacy icons, labels and notices to convey how personal data are handled~\cite{fox2018communicating,kelley2009nutrition}.

\textbf{Privacy Policy Visualization}: A privacy policy, also referred to as a privacy statement, is a legal document that describes how an
organization collects, processes, and shares its users' personal information, how the integrity
and confidentiality of this information is ensured, and what rights the users have to control
access to this data~\cite{PrivacyPolicies2002}. Within the broad spectrum of privacy visualization, we define \textit{Privacy Policy Visualization} as \textbf{the specific application of visual or multimodal design to showcase privacy policies--legal documents detailing data collection, processing and protection protocols~\cite{PrivacyPolicies2002}}. As shown in Figure~\ref{fig:scope}, privacy policy visualization constitutes a specialized sub-field nested within generic privacy visualization. 

We distinguish these concepts by their primary function: whereas generic privacy visualization often focuses on operational controls (e.g., toggle switches) of status indicators (e.g., camera LEDs), privacy policy visualization specifically addresses the informational transparency gap inherent in legal documentation. Its goal is interpretive--translating static, opaque legal text into accessible formats to support valid informed consent (e.g., Privacy Bird \cite{cranor2006user}, Privacy Nutrition Label \cite{kelley2009nutrition}). Figure~\ref{fig:scope} further highlights that this domain is dynamically interconnected with adjacent research streams: it is supported by \textbf{automatic generation technologies}, informs \textbf{large-scale analysis}, and is grounded in \textbf{user and developer understanding}. Notably, while consent interfaces~\cite{habib2022okay} may serve dual purposes as both a notice and a control, our scope is strictly limited to the visualization of the policy content itself.

We further delineate privacy policy visualization through distinguishing privacy notices from privacy controls. A privacy notice focuses on transparency, informing users about data practices, whereas a privacy control focuses on agency, allowing users to influence those practices. While early research treated these as separate entities, modern approaches frequently embed controls within the notice to facilitate informed decision-making~\cite{schaub2015design}. Therefore, we adopt this term as an umbrella covering both the static representation of policy text and the interactive interfaces derived from it. 

\subsection{Analytical Framework}\label{sec:framework}

To systematize the fragmented literature, we constructed a multi-dimensional framework adapted from interaction design lifecycles~\cite{salinas2020systematic, hasani2020user}. We selected this lifecycle as a heuristic lens to organize the literature into a coherent narrative, recognizing that privacy policy visualization is fundamentally an HCI challenge requiring iterative alignment between user needs and system constraints. Although the design lifecycle~\cite{salinas2020systematic,hasani2020user} could contain aspects like \textit{understand context}, \textit{define requirements}, \textit{ideation}, \textit{design}, \textit{prototype}, \textit{data collection}, \textit{build}, and \textit{evaluation} when situated in different scenarios, our framework (Figure~\ref{fig:teaser}) structures the analysis along four most common lifecycle aspects: 

$\bullet$ \textbf{Understand context of use}~\cite{dananjaya2024user,hasani2020user}: We analyze the problem space by coding for the diversity of \textit{Target Audiences} and \textit{Device platforms} (e.g., Mobile vs. IoT). 
The trajectory of these dimensions reveals that the context is governed by the \textbf{Tension between Generality and Specificity}, where visualization faces the conflict between deploying generalized solutions across devices and demographics, versus crafting specific solutions for specialized environments.

$\bullet$ \textbf{Establish requirements}~\cite{hasani2020user}: We examine the functional necessities by synthesizing the specific \textit{User Challenges} (e.g., cognitive overload) cited in the literature. This analysis identifies that a primary requirement for privacy policy visualization is to resolve the trade-off between \textit{Information Load and Decision Efficacy}, defining the boundary conditions for user attention.

$\bullet$ \textbf{Design solutions}~\cite{dananjaya2024user,rahman2022systematic,hasani2020user}: We categorize the construction of artifacts by tracking the co-occurrence of \textit{Visualization Metaphors}, \textit{Information Content} and \textit{Technology \& Automation Level}. Tracing these codes demonstrates a \textit{Co-evolutionary Dynamic of Design and Automation}, where novel designs require advanced NLP/LLM techniques, and conversely, new automated capabilities unlock new interaction paradigms.

$\bullet$ \textbf{Evaluation}~\cite{hasani2020user,ortiz2023assessing}: We assess validation strategies by coding for \textit{Evaluation Methodology} and reflecting on the core tensions during the deployment of the design. We interpret these evaluation types as proxies for distinct priorities, revealing that the success of visualization hinges on \textit{Balancing Stakeholder Opinions}, especially aligning the conflicting incentives of regulators, developers, and users.

\subsection{Literature Selection Methodology}\label{sec:methodology}

\subsubsection{Literature Search and Selection}

We conducted a systematic literature search in digital libraries including ACM and IEEE, following the PRISMA framework~\cite{page2021prisma}. Using the query (``privacy'') AND (``policy'') AND (``demonstration'' OR ``visualization'') and related synonyms, we targeted English-language publications since 2000. This search was informed by the research trends, advancements in the field (e.g., early adoptions such as P3P~\cite{cranor2002web}) and paper volume. After removing duplicates, 719 paper remained, which were screened the title and abstract to directly exclude those irrelevant papers, retaining 101 papers. 

\begin{figure*}[!htbp]
    \centering
    \subfloat[Publication numbers per year.]{
        \includegraphics[width=0.5\textwidth]{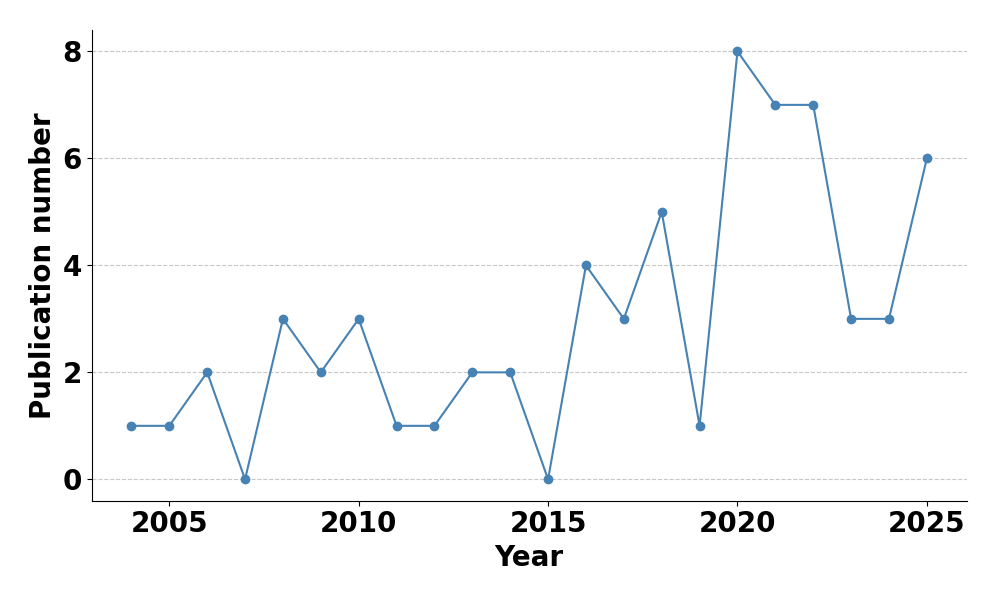}
         \label{fig:publication_number}
    }
    \subfloat[Publication number across venues.]{
        \includegraphics[width=0.5\textwidth]{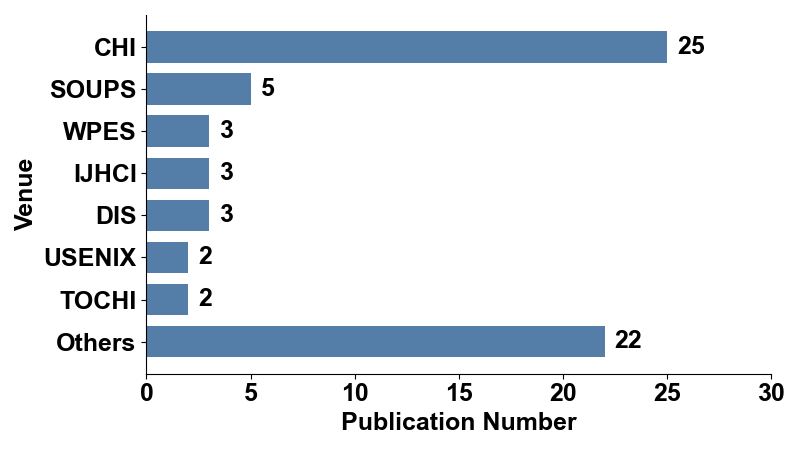}
        \label{fig:distribution}
    }
    \caption{The (a) publication numbers per year, and (b) distribution of papers across different venues.}
    \Description{Two statistical charts visualizing the dataset of the literature review. (a) The left chart is a line graph titled ``publication numbers per year'' with the x-axis ranging from 2004 to 2025 and the y-axis from 0 to 8. The data line shows a fluctuating upward trend, remaining low until 2015, the rising significantly to a peak of 7 papers annually between 2020 and 2023, before dipping to 5 papers in 2025. (b) The right chart is a horizontal bar chart titled ``publication number across venues''. The y-axis lists venues and the x-axis shows counts. ``CHI'' has a long bar with 25 publications, and ``others'' with 27. ``SOUPS'' has 5, while ``IJHCI'', ``DIS'' and ``TOCHI'' have shorter bars ranging between 2 and 3 publications.}
   \label{fig:number_venue}
\end{figure*}

A subsequent eligibility stage excluded works outside the scope of this paper with four criterion: (1) we exclude those with only \textit{theoretical or analytical contribution}, which focuses purely on theoretical frameworks of analyses without a concrete visualization design or system (N=8). (2) We exclude those \textit{focus on NLP}, which centers on NLP contributions for policy generation or automated analysis, rather than user-facing visualization component (N=19). (3) We exclude those focus on general privacy, which discussed general privacy topics (e.g., bystander privacy, privacy attitudes) not directly related to policy visualization (N=16). (4) We exclude those focus on policy itself, which analyzed the policy text (e.g., readability metrics) without designing or evaluating a visualization (N=6). This screening resulted in 52 relevant papers. To complement the search and ensure no omission of other papers, we performed forward and backward snowballing~\cite{wohlin2014guidelines,webster2002analyzing}, as well as consulting other researchers within the institution. These two processes added 9 and 4 papers each, totaling 13 papers, yielding a curated collection of 65 papers. Figure~\ref{fig:publication_number} shows the annual publication numbers, with the number increasing from 2016 to 2020, peaking from 2020 to 2023. Figure~\ref{fig:distribution} shows the publication numbers across venues, where CHI has the highest number of papers published. Further details on the search process and selection criteria are provided in Appendix~\ref{app:search_selection}. 

\subsubsection{Analysis and Synthesis}

The analysis was conducted with two stages, first performing a coding combining inductive and deductive analysis, and then an interpretative synthesis. For the first stage, our analytical framework was organized around the four phases of design (i.e., context, requirements, design, evaluation)~\cite{hasani2020user} to ensure comprehensive coverage, and the specific coding criteria were developed empirically. We adopted a hybrid coding strategy combining deductive and inductive approaches~\cite{fereday2006demonstrating}. We deductively established high-level dimensions (e.g., ``Format'', ``Challenge'') based on prior literature~\cite{mcdonald2022privacy,herriger2025context, obar2020biggest,reinhardt2021visual}. To instantiate these dimensions and explore other possible dimensions, four authors conducted a pilot coding on a representative subset of papers (N=10). Through open coding, we identified specific, data-driven categories (e.g,. filling the ``Format'' dimension with concrete codes like ``Nutrition Label'', ``Chatbot'' and ``Comic''). This iterative process produced the final codebook for the first stage (see Appendix~\ref{app:codebook}). The coding dimensions consisted four primary aspects: \textit{context of use}, \textit{requirements}, \textit{design solutions} and \textit{evaluation}. \textit{Context of use} consisted of \textit{target audience} and \textit{device \& platform context}. \textit{Requirements} consisted of \textit{problem definition} and \textit{contribution type}. \textit{Design solutions} consisted of \textit{visualization metaphor}, \textit{information content}, \textit{interaction timing}, \textit{design rationale} and \textit{automation level}. \textit{Evaluation} consisted of \textit{evaluation methodology} and \textit{success metrics}. The refined scheme was then applied to the full dataset. To ensure reliability, all the rest of the papers were coded by one author, and were checked by all the rest of the research team. Discrepancies were resolved through meetings and discussions.

Moving beyond descriptive statistics, we engaged in a recursive process of close reading and interpretive synthesis to derive at the four overarching findings. The aim was to identify latent tensions and historical trajectories within the corpus. We achieved this process through repetitively traversing through the papers, comparing between codes, generating candidate patterns, discussing with all research team members, and ground the findings in the papers. During this process, to arrive at design recommendations and implications, we focused on (1) juxtaposing the coded dimensions to uncover systemic conflicts, and (2) cluster papers based on their similar properties and tracing the temporal shifts. For example, we correlated \textit{Automation Levels} with the complexity of \textit{Visualization Metaphors}, and observed their temporal shifts to uncover the co-evolution of design and automation. This processes iteratively set the sub-themes, and further aggregated into themes (four patterns). One primary author led this process, and when new patterns or sub-themes emerged, the author discussed with all other authors in the research team to challenge the sub-themes and the themes. The themes and sub-themes are shown in Sections 4 to 7.

\subsection{Contributions}\label{sec:contribution}

Our review adopts temporal evolution as the lens to analyze privacy policy visualization, aiming at providing insights for future privacy policy visualization design: 

\textbf{[Synthesis] A systematic temporal synthesis:} We organize the literature along the core phases of the design lifecycle (Figure~\ref{fig:teaser}), enabling a structured examination of how the domain has matured.

\textbf{[Patterns] Identification of four key patterns in privacy policy evolution:} We identify four key patterns around the evolution of privacy policy visualizations, corresponding to context, user requirements, design and evaluation. These patterns allow for a systematic mapping of the developmental trajectory of privacy policy visualization. 

\textbf{[Implications] Actionable guidelines for the CHI community:} Drawing upon principles, we distill four sets of actionable guidelines to inform future research on privacy policy visualization, especially for the CHI community. 

\section{Related Works}

Prior literature on privacy policies primarily focuses on three distinct lenses: textual evolution~\cite{wagner2023privacy,amos2021privacy,belcheva2023understanding}, computational analysis~\cite{adhikari2023evolution,van2024privacy}, and user comprehension~\cite{raibulet2025awareness,mitchell2022privacy}. We review these streams to situate our contribution within the HCI landscape.

One primary stream analyzes the policy text. Longitudinal analyses indicate that policies have shifted toward greater length and reduced readability over time~\cite{wagner2023privacy,amos2021privacy}, while often exhibiting increased ambiguity~\cite{belcheva2023understanding}. Other studies have accessed compliance with regulatory frameworks like General Data Protection Regulation (GDPR)~\cite{linden2020privacy} or examined ecosystem-specific practices~\cite{hashmi2021longitudinal}. While these works effectively map the changing legal and textual landscape, they did not focus on the design and evolution of visual interfaces used to present this information.

A second body of work focuses on computational methods for policy analysis. This includes systematic reviews of NLP and machine learning techniques for information extraction~\cite{del2022systematic,adhikari2025natural,van2024privacy}, alongside surveys of machine-readable standards like P3P~\cite{kumaraguru2007survey,kasem2015security}. Research has also characterized the technical challenges inherent in these automated pipelines~\cite{mhaidli2023researchers}. These studies establish the technical feasibility of processing policy data. However, our review diverges by examining how such data is visually translated for end-users. 

Third, user-centric research has documented the challenges of policy comprehension. Studies suggest that current disclosure practices are often hard to understand~\cite{raibulet2025awareness} and contribute to the ``privacy paradox'', where protective behaviors lag behind stated concerns~\cite{mitchell2022privacy,barth2021data}. Related reviews have also examined informed consent within HCI research ethics~\cite{schwind2025scoping} or in general~\cite{verreydt2021security}. While this literature identifies the gap between users and policies, rather than identifying the problems, our work focuses on analyzing historical trajectory visualization to solve this problem.

The studies most relevant to our work involve either temporal analysis~\cite{adhikari2023evolution,wagner2023privacy} or design systematization~\cite{barth2022understanding}. For instance, Wagner et al.~\cite{wagner2023privacy} utilized NLP to track temporal shifts in data collection, yet their analysis remained limited to the textual domain. Similarly, Barth et al.~\cite{barth2022understanding} reviewed privacy visualization through a ``privacy by design'' lens but did not address the broad historical evolution of privacy policy visualization strategies. Our review addresses these by providing a systematic analysis of how privacy policy visualization evolved over time to provide insights to the HCI community.

\section{Finding 1: Tension Between Generality and Specificity}\label{sec:general_specific}

\begin{figure*}[!htbp]
    \includegraphics[width=\textwidth]{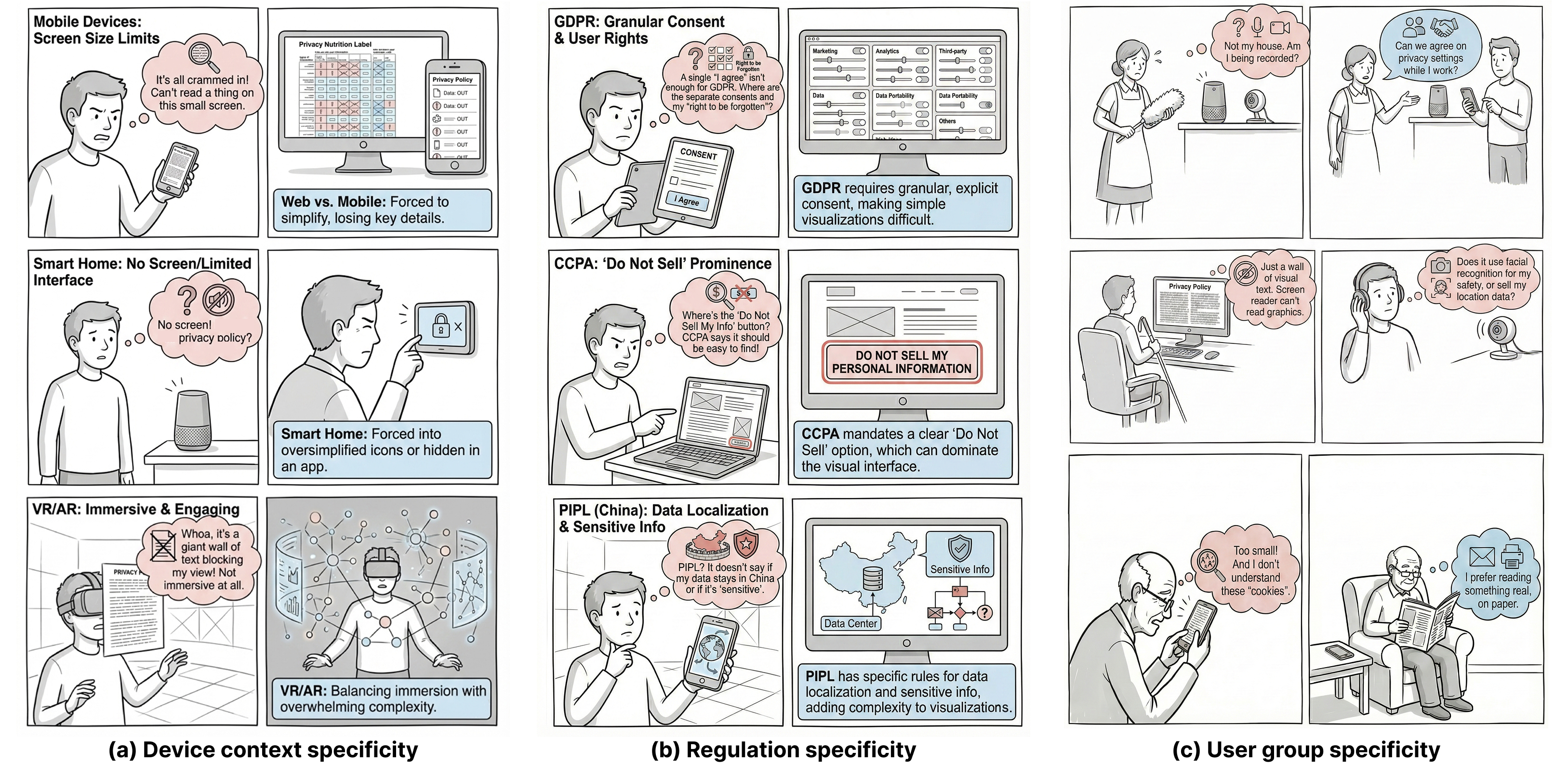}
    \caption{The illustration of the change from general solutions to specificity, (a) device contexts~\cite{lim2022mine,thakkar2022would}, (b) regulations~\cite{zhang2025privcaptcha}, (c) user groups~\cite{he2025privacy,sharma2025before}. (The pictures were generated using Gemini Nano Banana and edited by the authors.)}
    \label{fig:general_specificity}
    \Description{A three-panel comic-style illustration contrasting generic versus specific design contexts. (a) Device context specificity: The left column compares three scenarios: a user squinting at a mobile phone, with labels ``Screen Size Limits'', a user confused by a smart speaker, with labels ``No Screen/Limited Interface'', and a user in a VR headset blocked by a floating text document, with labels ``VR/AR: Immersive and Engaging''. (b) Regulation specificity: The middle column shows screens adapting to different laws, with a ``GDPR'' screen showing granular consent toggles, a ``CCPA'' screen highlighting a large ``Do Not Sell My Personal Information'' button, and a ``PIPL'' screen illustrating data localization maps and sensitive info diagrams. (c) User group specificity: The right column depicts diverse user needs, with a domestic helper interacting with smart home devices, a visually impaired user listening to audio privacy cues, and an elderly user reading a simplified tablet interface.}
\end{figure*}

Innovation in privacy policy visualization is driven by a tension between generality and specificity. On one hand, there is a strong design imperative for generalizable solutions--standardized formats that minimize cognitive load through familiarity and apply broadly across demographics and platforms. On the other hand, the emergence of distinct operational contexts and evolving technologies creates a competing demand for specificity, as generic models often fail to accommodate the unique affordances and constraints of new hardware. This dynamic interplay has forced the field to evolve from ``one-size-fits-all'' paradigms toward highly specialized solutions tailored to specific devices, regulatory frameworks, cultures, and stakeholder needs, as illustrated in Figure~\ref{fig:general_specificity}.

\subsection{\textcolor{color1}{Adaptation to Device Contexts}}

The research trajectory demonstrates a progression from platform-agnostic ideals toward device-specific implementations. Initial academic efforts prioritized universality, developing standardized visualizations for the web such as P3P~\cite{cranor2002web,cranor2006user} and privacy labels~\cite{kelley2010standardizing,kelley2009designing}. However, the proliferation of diverse hardware ecosystems and device types soon mandated more specialized approaches. For instance, the limited screen spatial constrains of mobile devices render the verbose tabular impractical. Researchers responded by proposing techniques to adapt for the interface recognition features for mobile devices, such as OCR-based automation~\cite{pan2024new}. To adapt for the screen space, Kelly et al.~\cite{kelley2013privacy} proposed interface adaptations like ``privacy facts'' which is displayed in app stores. To address the opacity of mobile data flows, subsequent work focused on using visual depictions, scores, and monetary illustrations for depicting collected data and risks~\cite{stover2021investigating}. There were also work devising mobile dashboards for sensing data control and deletion~\cite{bemmann2022influence}. As mobile ecosystems matured, designs became increasingly granular, incorporating user rankings on the app stores~\cite{shiri2024motivating}, and specialized tools like AppAware~\cite{paspatis2020appaware} to track permissions. Recent innovations continue to address these spatial and temporal boundaries through adaptive interfaces such as CAPTCHA-based interactions~\cite{zhang2025privcaptcha}.

Parallel to the mobile revolution, the ubiquitous and screenless nature of the IoT necessitated a departure from traditional display-basd policies. Unlike web services requiring active consent, IoT data collection is often passive, inconspicuous, and pervasive~\cite{fruchter2018consumer,steinberg2014these}, posing risks to multiple users in shared spaces~\cite{ziegeldorf2014privacy,naeini2017privacy,zheng2018user}. Since IoT devices lack canonical spaces for visualization and involve vague connection topologies~\cite{heino2023assessment}, researchers developed bespoke communication channels. Bahrini et al.~\cite{bahrini2022s} explored companion apps as a visualization medium, comparing list-based and device-based layouts. To structure this complexity, Feng et al.~\cite{feng2021design} introduced a specialized framework for smart homes, which organizes practices by type, functionality, and timing. They further instantiated this framework as an IoT Assistant (IoTA). Other work has focused on specific IoT challenges. For instance, Thakkar et al.~\cite{thakkar2022would} investigated dashboards to reduce users' and bystanders' dependence. Beyond privacy policy visualization, Roy et al.~\cite{roy2025privacyvis} visualized sensor workflows to expose processing risks, and Muhander et al.~\cite{al2023interactive} defined key privacy factors (e.g., storage, retention) to guide the creation of effective, domain-specific notices.

Most recently, the rise of immersive platforms has propelled a reimagining of privacy policy visualization beyond text. In Virtual Reality (VR), traditional textual policies suffer from poor readability and break immersion. Consequently, the paradigm may shift from passive reading to active, embodied experience. Lim et al.~\cite{lim2022mine} proposed role-play-based solutions to enhance comprehension through engagement, while Stellmacher et al.~\cite{stellmacher2022escaping} demonstrated the efficacy of gamified, immersive approaches like Escape Games for privacy education. Beyond privacy policy visualization, Windl et al.~\cite{windl2025designing} also characterized the design space for consent mechanisms, emphasizing that immersive environments require targeted visualization mechanisms.

\subsection{\textcolor{color1}{Response to Legal Frameworks}}

The drive toward specificity in privacy research is heavily influenced by external regulatory mandates. As distinct legal frameworks impose varied data handling requirements, visualization research has increasingly tailored its design goals to ensure compliance with specific legislation. Feng et al.~\cite{feng2021design} note that while regulations such as the GDPR and the CCPA introduce rigorous requirements for privacy choices, these high-level legal concepts require HCI research to translate them into actionable guidance for practitioners. Similarly, Srinath et al.~\cite{srinath2021privaseer} highlight that these laws fundamentally assume users will read and understand policies before providing consent, a premise that drives the need for more effective automated retrieval and presentation systems.

This legal pressure has led to the development of visualizations and tools targeting specific compliance mechanisms, particularly regarding user consent and opt-out rights. Mohammadi et al.~\cite{mohammadi2019pattern} focused on the GDPR's requirements for informed consent and purpose specification, aiming to make end-users aware of their rights regarding data processing. Addressing the ``notice and choice'' requirements common to both GDPR and CCPA, Kumar et al.~\cite{bannihatti2020finding} developed techniques specifically to identify opt-out provisions. In a more targeted design intervention, Habib et al.~\cite{habib2021toggles} evaluated icons and textual descriptions for the CCPA-mandated ``Do Not Sell My Personal Information'' right, finding that specific link texts were necessary to accurately communicate these legal options to users.

Beyond Western frameworks, the scope of research has expanded to accommodate regional and country-specific regulations. Zhang et al.~\cite{zhang2025privcaptcha} based their visualization design choices on the Personal Information Protection Law (PIPL) in China, while Zhao et al.~\cite{zhao2022fine} constructed the CA4P-483 dataset to address the Cybersecurity Law of the People's Republic of China (CLPRC) alongside GDPR and CCPA. On a broader scale, Javed et al.~\cite{javed2024systematic} emphasize that privacy systems must account for a diverse landscape of global standards, including the Organisation for Economic Co-operation and Development (OECD) guidelines, Fair Information Practice Principles (FIPPs) and regional acts such as the Electronic Communications and Transaction Act (ECTA) and Personal Data Protection Act (PDPA). This legislative diversity necessitates design solutions that are not only user-centric but also adaptable to varying legal jurisdictions.

\subsection{\textcolor{color1}{Addressing Specific User Groups}}

While general-purpose privacy policy visualization aim to broadly reconcile the information asymmetry between service providers and end-users~\cite{mohammadi2019pattern}, they often assume a uniform user base. Early innovations, such as the expandable grid concepts by Reeder et al.~\cite{reeder2008expandable}, focused on creating standardized interfaces usable by a general audience. However, this ``one-size-fits-all'' approach may not fit for the nuanced needs of users situated in specific social contexts. For instance, in smart home environments, Thakkar et al.~\cite{thakkar2022would} demonstrated that primary users and bystanders differ significantly in their preferences for privacy notifications. Similarly, recent research has moved beyond the individual user model, exploring group-based designs for app protection awareness~\cite{shiri2024motivating} and collaborative frameworks that leverage family members to support informed decision-making~\cite{aljallad2019designing}.

Beyond social context, standard visualizations may not accommodate the accessibility requirements of specific populations, particularly the elderly, children, and Blind of Low Vision (BLV) individuals~\cite{chen2025clear,zhang2025privcaptcha}. For BLV users, purely visual policy presentations are insufficient. Feng et al.~\cite{feng2024understanding} highlighted that BLV users possess distinct risk perceptions and information behaviors that necessitate non-visual interventions. Furthermore, Sharma et al.~\cite{sharma2025before} explored the use of Generative AI to interpret visual content for this demographic, suggesting that BLV users often balance efficiency, privacy and emotional agency. These findings underscore the potential of developing auditory or multimodal privacy presentations that align with the specific sensing capabilities of BLV users.

Age-specific differences in mental models further challenge the efficacy of general privacy visualizations. Oates et al.~\cite{oates2018turtles} found that children conceptualize privacy through physical boundaries--such as bedrooms or bathrooms--rather than abstract data practices, creating a disconnect with standard policy terminology. Zhang et al.~\cite{zhang2025privcaptcha} also highlighted how their techniques may fit for child-specific regulations and corresponding privacy policy text. Aly et al.~\cite{aly2024tailoring} note that most existing support interventions are designed for general populations, which may not be suitable for the elderly. These work suggests that empowering older adults requires specific educational modalities that match their digital literacy levels and learning styles.

\subsection{\textcolor{color1}{Future Directions: Towards Customized Privacy Policy Visualization}}

Faced with these tensions, future research should move beyond simply creating specific tools and focus on bridging generality and specificity. We therefore propose three directions to bridge specific contexts with unified, actionable protection: \textbf{adaptive generative interfaces}, \textbf{context-appropriate situated visualization},  and \textbf{integrated visualization mechanisms}.

\textbf{Adaptive Generative Interfaces.} Current research addresses specific user groups through manually designed adaptations~\cite{feng2024understanding}. Future work could leverage generative and malleable interfaces~\cite{min2025malleable,shaikh2025creating} to automate this specificity. By decoupling the semantic content of privacy policies from their presentation and introducing intermediary representations~\cite{zhang2025privcaptcha}, systems can utilize LLMs to dynamically render interfaces tailored to the user's real-time states and device constraints. For example, the same underlying policy metadata could be instantaneously generated as a gamified illustration for a child~\cite{oates2018turtles}, or a high-contrast audio summary for a BLV user~\cite{feng2024understanding}. This shifts the design paradigm from designing separate tools for specific groups for creating intelligent systems that autonomously adapt to user demographics.

\textbf{Context-Appropriate Situated Visualization.} To address the hardware constraints of screenless and immersive platforms without imposing user burdens, visualizations could be designed to align the device's supported modality~\cite{thakkar2022would}. In standard smart home contexts, research should prioritize ambient peripheral displays, such as subtle lighting shifts or haptic feedback, that signal data transmission without demanding focused attention~\cite{bahrini2022s}. Conversely, for users explicitly operating within VR, systems could develop in-situ overlays that visualize sensor coverage directly in the virtual environment~\cite{zhao2025arena}, such as rendering invisible camera fields of view~\cite{lim2022mine}. By selecting the visualization channel that matches the user's current engagement, designers can resolve the invisibility of ubiquitous computing without forcing users to adopt cumbersome behaviors.

\textbf{Integrated Visualization Mechanisms.} To resolve the fragmentation of devices and regulatory context, future research could explore decentralized interoperability standard~\cite{troncoso2017systematizing}, where privacy metadata is synchronized across all user devices~\cite{zhou2024bring}. This would enable a multi-device visualization hub, allowing a user to inspect and control data flows from a smart appliance directly via a mobile phone or desktop dashboard~\cite{thakkar2022would}. To address the disconnect between abstract legal texts (e.g., GDPR, PIPL) and concrete user risks, future tools could employ sandbox-based counterfactual simulations~\cite{chen2024empathy}. Instead of static text describing potential risks, systems should utilize agents to simulate and visualize the specific consequences of a permission grant before the user consents, such as demonstrating how disclosed location data could be reconstructed by an adversary.

\section{Finding 2: Trade-off Between Information Load and Decision Efficacy}\label{sec:load_efficiency}

\begin{figure*}[!htbp]
    \centering
    \includegraphics[width=\textwidth]{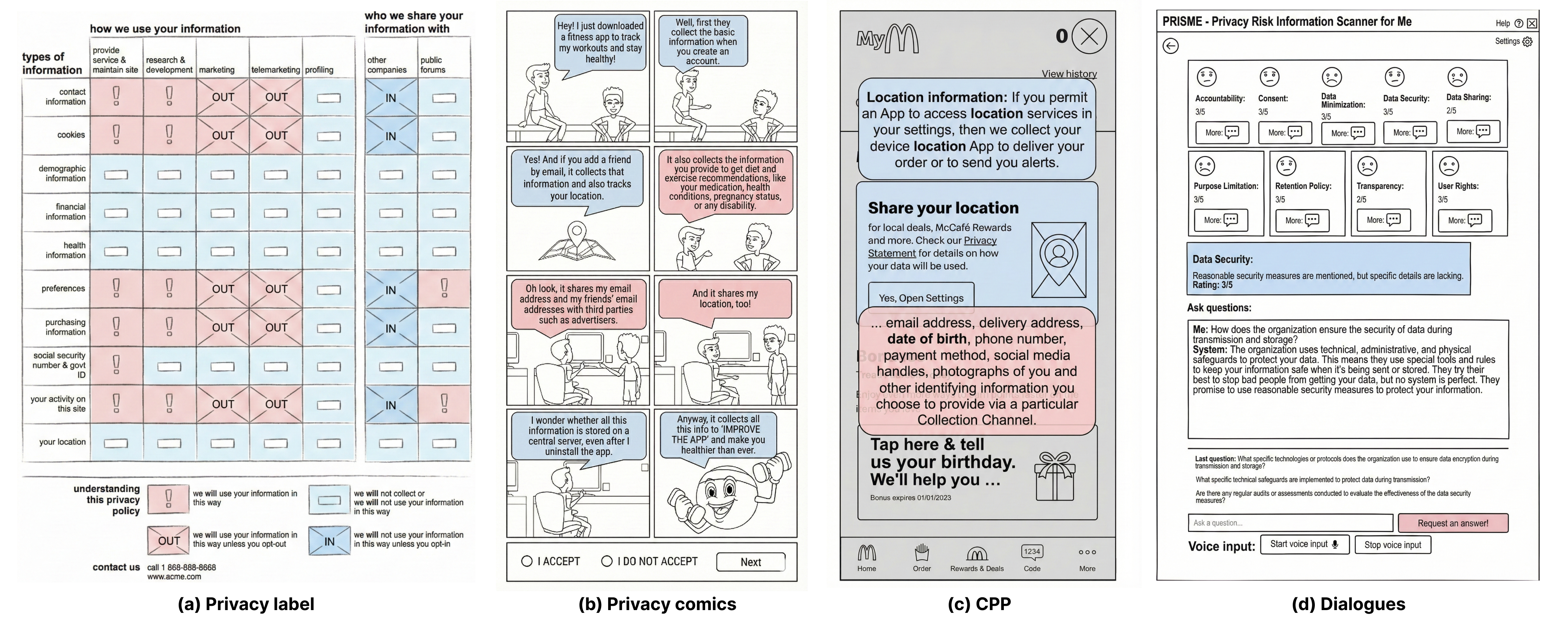}
    \caption{Illustration of the representative privacy policy visualization for Finding 2: (a) Privacy label~\cite{kelley2009nutrition} for the paradigm of \textit{succinct and structured summarization}, (b) Comics~\cite{tabassum2018increasing} for the paradigm of \textit{enriched visuals and narratives}, (c) CPP~\cite{pan2024new} for the paradigm of \textit{contextual integration to balance load}, (d) Dialogues~\cite{freiberger2025you} for the paradigm of \textit{proactive and interactive support}. (The pictures were generated using Gemini Nano Banana and edited by the authors.)}
    \label{fig:illustration_design2}
    \Description{Four distinct interface mockups illustrating different privacy policy visualization metaphors. (a) Privacy label: A tabular grid resembling a nutrition label. Rows list data types, such as contact info and health, and columns indicate usage, such as used for tracking, linked to user, with ``Opt-In or Opt-Out'' toggle switches. (b) Privacy comics: A four-panel comic strip where illustrated characters discuss data collection practices in speech bubbles, providing a narrative explanation. (c) Contextual Privacy Policy: A wireframe of a mobile app screen. Pop-up labels are attached to specific UI elements. For example, a bubble pointing to a map explains ``Location Information: If you permit, we collect your device location.'' (d) Dialogues: A chat interface titled ``Privacy Risk Information Session.'' It shows a conversation history where the user asks ``How is my data used?'' and ``What encryption?'', and the system responds with specific text bubbles explaining service improvement and AES-256 encryption standards.}
\end{figure*}

We found with the increase of privacy policy information presented, user's ability to make effective decisions decreased. Excessive information load introduces cognitive fatigue, leading to resignation and rendering the communicative intent of privacy policies ineffective. This phenomenon underscores the failure of the traditional ``notice-and-choice'' model, which depends on the flawed assumption that users can rationally evaluate lengthy legalistic texts~\cite{cranor2012necessary,staff2011protecting,nissenbaum2011contextual}. Consequently, the research community has shifted focus from merely presenting information to restructuring how privacy concepts are communicated. We categorize this evolution into four paradigms aimed at enhancing users' decision efficacy (Figure~\ref{fig:illustration_design2}): (1) \textbf{Succinct and structured summarization}, which condenses text into standardized formats, (2) \textbf{Enriched visuals and narratives}, which leverages graphical and storytelling elements to enhance comprehension, (3) \textbf{Contextual integration to balance load}, which balances information load by distributing information disclosure to relevant moments of interaction, and (4) \textbf{Proactive and interactive support}, which transitions users from passive reading to active, tool-assisted exploration.

% \begin{figure}[!htbp]
%     \includegraphics[width=0.75\textwidth]{Figure/load-efficacy.png}
%     \caption{The illustration of the relationship between cognitive load and decision efficacy from the earliest phase to the current landscape.}
%     \label{fig:load_efficacy}
% \end{figure}

\subsection{\textcolor{color2}{Succinct and Structured Summarization}}

The initial response to information overload focused on reducing density through standardization and abstraction. Recognizing that the legalistic nature of policies impedes comprehension~\cite{nissenbaum2011contextual}, early methodologies aimed to extract and present core data practices in user-friendly, layered interfaces~\cite{cranor2006user}. A notable development in this domain was the privacy ``nutrition label'', which standardized key attributes--such as data recipients, categories, and retention--into a tabular, at-a-glance format aimed at facilitating rapid comparison~\cite{kelley2009nutrition,kelley2010standardizing}. Theoretical frameworks like the Privacy Policy Visualization Model (PPVM) further formalized this approach by defining essential policy dimensions (e.g., visibility, granularity) to guide structured representations~\cite{ghazinour2009model}.

Within this paradigm, researchers navigated the tension between abstraction and precision. For instance, Expandable Grids~\cite{brodie2006empirical,lipford2010visual} offered a compact, high-level overview with ``drill-down'' capabilities, prioritizing generalizability and quick settings modification. In contrast, interfaces like AudienceView provided concrete visualizations of social connections to enhance contextual clarity, albeit at the cost of scalability and increased navigational effort. While these structured approaches successfully reduced initial information load, they often failed to fundamentally motivate users to engage with the underlying policy text.

\subsection{\textcolor{color2}{Enriched Visuals and Narratives}}

To address the lack of user motivation, subsequent research explored transforming static text into visually immersive and narrative formats. Moving beyond tabular layouts, these methods leveraged graphical design and storytelling to lower the barrier to entry. Techniques ranged from applying typographic design principles to highlight critical End User License Agreements (EULA) clauses~\cite{kay2010textured} to employing comic-based narratives than explain data practices in vivid, relatable scenarios~\cite{tabassum2018increasing}. Other similar methods also used flyers~\cite{kulyk2019does} or paraphrased clauses~\cite{waddell2016make} to enhance readability. More recent explorations proposed to let the social family to help the user in making informed privacy decision~\cite{aljallad2019designing}. Although it was effective, they found people may lack the motivation to reach out and help others. Other work have extended this into immersive domains. For example, theatrical installations in VR environments have been used to provoke reflective understanding of privacy risks through artistic engagement~\cite{jung2023art}.

Beyond engagement, visualization has also been employed to elucidate complex data flows. Privacy Wheel used wheel-like interfaces to display key privacy concepts~\cite{van2012happens}. Tools like PrivacyInsight utilized graph-based visualizations to trace the lineage of personal data, allowing users to isolate and follow the trajectory of specific data points~\cite{bier2016privacyinsight}. Poli-see built upon Data Track, using a graphical representation to address data type, usage, transformation, and available configuration options~\cite{guo2020poli}. Similarly, the Privacy Policy Permission Model (PPPM) captured policy logic as diagrams to provide a uniform representation of data usage~\cite{majedi2024privacy}. While these visually enriched formats substantially improved comprehensibility and engagement in experimental settings, their ecological validity remains a challenge, as users in real-world deployment often lack the intrinsic motivation to interact with complex visualizations solely for privacy management.

\subsection{\textcolor{color2}{Contextual Integration to Balance Load}}

Recognizing that \textit{when} information is presented is as critical as \textit{how} for users' decision efficacy, a third body of work focuses on contextuality and saliency. Schaub et al.~\cite{schaub2015design} delineated a design space emphasizing timing, channel, modality and control. Ebert et al.~\cite{ebert2021bolder} further empirically found that exclusive privacy label results in longer reading time and engagement than an embedded privacy label. These results manifests in the use of privacy icons and indicators embedded directly within user interfaces, serving as salient, intuitive reminders~\cite{gerl2018extending,harkous2018polisis}.

This paradigm also seeks to distribute the information load over time. By decomposing single privacy policy into contextual excerpts, systems can utilize users' cognitive resources more efficiently, presenting only relevant information at the point of decision-making. Research therefore proposed to reduce the number of contextual privacy decisions they have to make altogether~\cite{chitkara2017does}, thereby potentially reducing the information load. This evolved into the concept of Contextual Privacy Policies (CPPs)~\cite{ortloff2018evaluation}, which aims to align information disclosure with specific interaction events. However, even with improved timing, the effectiveness of these cues relies heavily on their ability to compete for user attention against the primary tasks they are performing~\cite{windl2022automating,ortloff2020implementation}. 

\subsection{\textcolor{color2}{Proactive and Interactive Support}}

The most recent evolution shifts the user's role from a passive reader to an active inquirer, empowered by interactive tools and automation. This paradigm posits that enhancing decision efficacy requires providing users with agency and proactive support. Early implementations, such as Privacy Policy Options~\cite{mohammadi2019pattern} introduced granular control panels, and Schufrin et al. developed a web-based tool that allows users to interactively explore the temporal aspect of their collected data~\cite{schufrin2020visualization}. Latter tools like interactive visual privacy policies~\cite{reinhardt2021visual} allowed users to manipulate opt-in/opt-out settings for specific data types dynamically. 

Currently, this domain is being reshaped by LLMs and conversational agents, which offer on-demand interpretation of policies. Systems like Pribots pioneered chat-based policy question and answering (Q\&A)~\cite{harkous2016pribots}, a concept recently advanced by LLM-driven solutions. For instance, PrivCAPTCHA integrates interactive CAPTCHA-based policy learning into smartphone usage~\cite{zhang2025privcaptcha}, while other works utilize LLMs to generate context-aware, risk-based warnings~\cite{chen2025clear}, or facilitate direct dialogue with policy texts~\cite{freiberger2025you}. These proactive agents represent a shift that rather than simplifying the document, they act as intermediaries that synthesize information and execute control, thereby keeping the user in the lop while minimizing the user's load of information processing~\cite{zhang2025towards}.

\subsection{\textcolor{color2}{Future Directions: Navigating the Cognitive Barrier with Proactive Techniques}}

% TO YING & EVE & YUTING:
% 未来的技术应该更多考虑现在没解决的问题，比如说形式上是不是还不够主动服务，主动服务是不是就能让用户避免懒得点的问题。主动服务是不是会有一些bias，比如说总是给用户推荐最敏感的东西，用户就不看不敏感的了，是否会造成类似信息茧房的效应，怎么避免。主动服务真的是准确的吗，如果有不准确的问题怎么避免。

The core challenge facing current privacy policy visualization is user neglect and habituation~\cite{zhang2025privcaptcha,reinhardt2021visual,chen2025clear}. To solve this long-standing challenge, future directions should consider dynamic intervention and delegated execution while remaining vigilant regarding the potential risks caused by proactivity.

\textbf{Countering Habituation with Dynamic Intervention.} To prevent users from blindly clicking through visualization, systems could move beyond static displays to context-aware interventions that create friction~\cite{windl2022automating} and stimulate reflective thinking~\cite{terpstra2021think}. Instead of overwhelming users with data, systems could anticipate high-risk scenarios and adapt to user literacy levels~\cite{dupree2016privacy}, and intervene when a user's action deviates from privacy norms, thereby acting as a contextual ``privacy nudge''~\cite{wang2013privacy}. Furthermore, since users often lack the domain knowledge to scrutinize policies effectively, systems should proactively recommend relevant questions or assessments~\cite{freiberger2025you}, guiding users to uncover risks they would otherwise overlook.

\textbf{From Transparency to Delegated Execution.} Transparency is futile without the power to act. Future systems should ``close the loop'' by coupling awareness with control, as demonstrated by Reinhardt et al.~\cite{reinhardt2021visual}. Building on AI's ability to analyze policies~\cite{harkous2016pribots,ahmad2020policyqa} and detect manipulative ``dark patterns''~\cite{mathur2019dark}, personal AI agents~\cite{morel2025ai} could not only flag risks, dark patterns and non-compliance issues, but automatically negotiate behaviors, configure settings or even execute data deletion requests~\cite{angulo2015usable,bier2016privacyinsight} on the user's behalf, making privacy protection the default state rather than an effortful behavior for users.

\textbf{Mitigating the Risks of Proactivity.} While automation reduces cognitive load, it introduces critical vulnerabilities regarding overtrust and bias. Users conditioned to rely on proactive agents may fail into ``blind acceptance''~\cite{ma2024effect} or place excessive trust in system warnings~\cite{kraus2020effects}. Besides, if an imperfect agent fails to flag a risk, users may assume privacy and share data carelessly. Additionally, there is a risk of creating an ``information cocoon''~\cite{sunstein2006infotopia}, where the agent surfaces only high-sensitive or relevant risks, potentially blinding the user to other threats. Future research should focus on designing robust interactions that maintain user vigilance and prevent the automation from becoming another driver for ignorance.

\section{Finding 3: Co-evolutionary Dynamic of Design and Automation}\label{sec:design_automation}

\begin{figure*}[!htbp]
    \includegraphics[width=\textwidth]{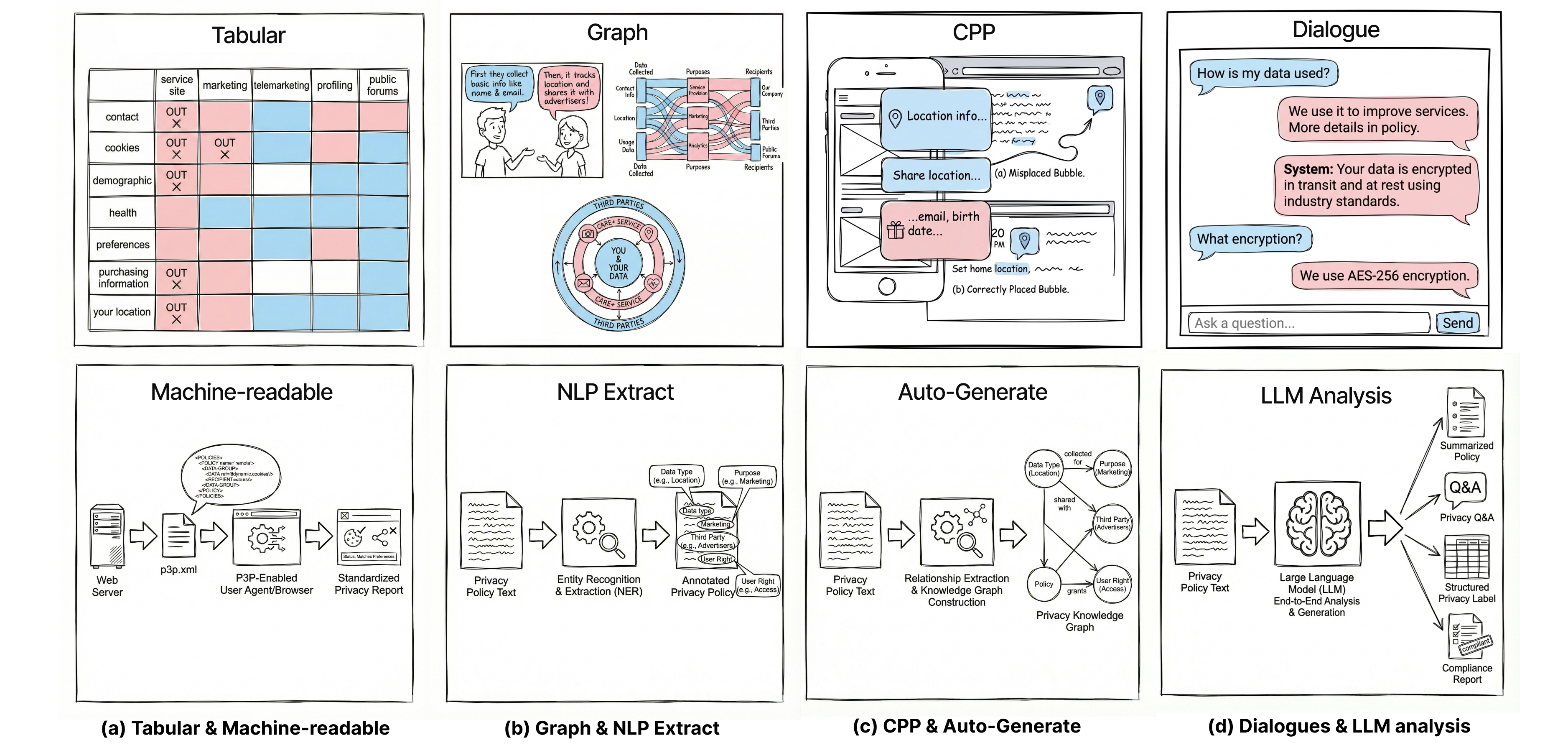}
    \caption{The co-evolutionary dynamic of design and automation, (a) \textit{Tabulars} and \textit{Machine-readable formats}, corresponding to the aspect of \textit{structured protocols and succinct forms}, (b) \textit{Graphs} and \textit{NLP extracting} techniques, corresponding to the aspect of \textit{graph-based narratives and automated extraction}, (c) \textit{CPPs} and \textit{Automatic analysis}, corresponding to the aspect of \textit{contextual privacy policy and its generation}, (d) \textit{Dialogues} and \textit{LLM-based analysis}, corresponding to the aspect of \textit{conversational interfaces and LLMs}. (The pictures were generated using Gemini Nano Banana and edited by the authors.)}
    \label{fig:design_automation}
    \Description{A 2 times 4 grid diagram illustrating the co-evolution of design metaphors on the top row, and their underlying automation technologies on the bottom row. Column 1 shows a ``Tabular'' grid design above, paired with ``Machine-readable'' technology below, which is illustrated as a server processing P3P or XML data. Column 2 shows a ``Graph'' visualization above, with sankey diagram and wheels, paired with ``NLP Extract'' technology below, illustrated as a document passing through an entity recognition process. Column 3 shows a ``CPP'' mobile overlay design above, paired with ``Auto-Generate'' technology below, illustrated as a knowledge graph construction pipeline. Column 4 shows a ``Dialogue'' chat interface above, paired with ``LLM Analysis'' technology below, illustrated as a brain icon processing policy text into question and answering, and summaries.}
\end{figure*}

As shown in Figure~\ref{fig:design_automation}, the evolution of privacy policy visualization is characterized by a co-evolutionary dynamic between design and automation. The pursuit of effective, user-centric design creates a demand for advanced automation technologies to instantiate and scale these conceptual frameworks. Conversely, breakthroughs in automation~\cite{harkous2018polisis,zaeem2018privacycheck} unlock new design possibilities, enabling the management of information complexity at previously unachievable scales. This feedback loop drives the transformation of privacy policies from static documents into intelligent, scalable services~\cite{chen2025clear,freiberger2025you}.

\subsection{\textcolor{color3}{Structured Protocols and Succinct Forms}}

Early efforts focused on machine-readable standards to facilitate automated processing. A prominent example is the P3P project by the World Wide Web Consortium~\cite{reagle1999platform}. This standardization enabled the creation of early user agents, such as Privacy Bird, which automatically processed structured policies to provide visual indicators to users~\cite{cranor2006user}. and prototypes like Privacy Finder, which integrated privacy policy summaries into search engine results~\cite{tsai2011effect,mcdonald2009comparative}. These technical initiatives highlighted the necessity for user-centric visualizations to improve human readability. In response to that, researchers proposed standardized formats analogous to nutrition labels. Kelley et al.~\cite{kelley2009designing} designed and validated the ``privacy nutrition label'', and standardized the visualization form of privacy labels~\cite{kelley2009nutrition,kelley2010standardizing}. Similarly, Reeder et al.~\cite{reeder2008expandable} explored expandable matrix interfaces to display policies, while Reinhardt et al.~\cite{reinhardt2021visual} later proposed visual interactive privacy policies that combined labels with opt-out controls.

\subsection{\textcolor{color3}{Graph-based Narratives and Automated Extraction}}

Following the standardized privacy labels approach, automation and design intertwined to foster each other. While these privacy label designs improved comprehension, manually populating them proved unscalable. This necessitated the development of automated information extraction techniques. Zimmeck combined machine learning classifiers with crowdsourcing to provide in-place policy assessments~\cite{zimmeck2014privee}. Subsequent research focused on specific clause and data practice extraction: Sathyendra et al.~\cite{sathyendra2017identifying} developed a two-stage classification pipeline to extract data collection practices and opt-out choices, while Bannihatti et al.~\cite{bannihatti2020finding} utilized BERT models for similar opt-out provisions. Tools like PrivacyCheck and PrivacyCheckv2~\cite{nokhbeh2020privacycheck,zaeem2018privacycheck} automated privacy policy summarization by answering questions regarding GDPR compliance and user control, incorporating competitors' privacy policy analysis. Furthermore, PI-Extract~\cite{bui2021automated} offered a fully automated system and a large-scale dataset for extracting data practices from privacy policies, and TLDR~\cite{alabduljabbar2021tldr} leveraged an ensemble of classifier to categorize data collection categories and detect missing information.

As extracting structured data became more feasible, design paradigms shifted toward more engaging visual narratives. Researchers explored novel presentations such as Sankey Diagrams~\cite{reinhardt2021visual}, privacy comics~\cite{tabassum2018increasing}, wheel-like interfaces~\cite{van2012happens}, graph-based visualizations~\cite{bier2016privacyinsight}, to make policies accessible to lay users. Poli-see also utilized a graphical representation to address data type, usage, transformation, and available configuration options~\cite{guo2020poli}, acknowledging automatic analyzing as a necessary step for scalability~\cite{wilson2018analyzing}. Similarly, PPPM captured policy logic as diagrams to provide a uniform representation of data usage~\cite{majedi2024privacy}.

Supporting these complex visualizations required automation capable of understanding relationships within the text, rather than isolated data points. Ravichander et al.~\cite{ravichander2021breaking} pointed out that NLP advances already facilitated tasks such as data practice identification, opt-out identification, compliance analysis, privacy question-answering, policy summarization and readability analysis.
To support relational visualizations, Cui et al.~\cite{cui2023poligraph} introduced PoliGraph, a knowledge graph approach that models policy statements as relations, developing the PoliGraph-ER system to automatically extract these structures from raw text.

\subsection{\textcolor{color3}{Contextual Privacy Policy and Its Generation}}

With the support of automatic extraction methods~\cite{sathyendra2017identifying,bui2021automated,nokhbeh2020privacycheck} and context-aware techniques, a significant milestone in this co-evolution was the shift toward CPPs. The core design philosophy of CPPs is that privacy information is most effective when presented in-situ, at the specific moment of a data practice~\cite{feth2017transparency,ortloff2020implementation,ortloff2018evaluation}. 

However, the scalability of CPPs was limited by the need to manually curate relevant text for every potential context. This bottleneck spurred research into automated text segmentation and generation. Tools like PrivacyInjector were developed to automate the delivery of context-aware visualizations in production environments~\cite{windl2022automating}. Subsequent research has focused on refining these automation frameworks to improve the accuracy and coverage of the generated snippets~\cite{pan2024new}, ensuring that the right information reaches the user at the right time. Recently, the advent of LLMs has further facilitated contextual displays, such as risk-based reflective notices based on LLM assessments~\cite{chen2025clear}.

\subsection{\textcolor{color3}{Conversational Interfaces and LLMs}}

The most recent evolutionary turn focuses on conversational interaction. Early work, such as PriBots~\cite{harkous2016pribots}, proposed chatbots to deliver policy content interactively. To overcome the scalability limits of short notices, Harkous et al.~\cite{harkous2018polisis} developed Polisis, which used privacy-specific embeddings to enable dynamic queries on privacy policy content. Recognizing the limitations of traditional retrieval, researchers developed datasets like PolicyQA~\cite{ahmad2020policyqa} to facilitate NLP-based privacy policy comprehension, though early BERT-based models still lagged behind human performance~\cite{ravichander2019question}. 

The advent of LLMs has fundamentally altered this landscape. Unlike traditional NLP, which often requires labor-intensive annotation~\cite{rodriguez2024large}, LLMs offer robust capabilities for summarization and question answering. Tools like Privacify~\cite{woodring2024enhancing} utilize LLMs for compliance-based summarization and multi-document analysis. While reliability remains a concern--prompting the development of evaluation benchmarks like GenAIPABench~\cite{hamid2024genaipabench}--recent studies indicate that LLMs can match or exceed traditional methods in extracting key policy aspects~\cite{rodriguez2024large}, and answering complex user queries~\cite{hamid2024genaipabench}.

This leap in automation capability is currently enabling a new generation of designs. LLMs are now being used for risk assessments and enabling dynamic QAs~\cite{freiberger2025you}, generate risk-based assessment notices~\cite{chen2025clear}, create intermediate representations and adaptive interfaces~\cite{zhang2025privcaptcha}, and thereby enabling these applications like CAPTCHA-based notifications~\cite{zhang2025privcaptcha}. These systems move beyond static summaries, leveraging the reasoning capabilities of LLMs to provide interactive privacy guidance, potentially empowering users with better informed consent, and spurring new end-to-end NLP tasks.

\subsection{\textcolor{color3}{Future Directions: Evolving the LLM-Based Co-evolutionary Relationship of Design and Automation}}

% TO YING & EVE & YUTING
% 未来会用LLM进一步实现原来的这些方式的端到端转换，例如说可能comic未来可以一键生成了，或许隐私教育和沟通可以有一些新范式的突破。并且LLM可以enable更快的prototype，去尝试更多的visualization form，甚至在早期帮助designer进行ideation，能让designer有更多的点子。但是当前时代也有一些新问题，比如LLM的hallucination, factual error之类的。

The integration of LLMs marks a paradigm shift towards generative interpretation, which synthesizes explanations and interfaces. This evolution offers the potential to bridge the long-standing gap between rigid legal disclosures and fluid user needs. However, it also introduces new risks regarding fidelity and agency. We identify three directions for future research to address these challenges and advance the field. 

\textbf{Advancing Design Through Generative Interaction.} Prior approaches relied on labor-intensive, manual creation of visualizations, such as privacy comics~\cite{tabassum2018increasing,knijnenburg2016comics},  games~\cite{stellmacher2022escaping} or VR scenarios~\cite{lim2022mine}. Future work could leverage LLMs to transform policy text to these intuitive modalities in an end-to-end manner. Beyond the final presentation, LLMs can serve as ideation partners for designers, rapidly prototyping visualization forms, thereby effectively converging the design and automation phases~\cite{liu2024ai}. 

\textbf{Contextual Risk Communications.} Existing systems largely focus on extracting policy segments~\cite{harkous2018polisis} and defining contextual snippets~\cite{pan2024new}. LLMs empower the interpretation of the implication of these segments to model user-perceived risk~\cite{acquisti2017nudges}, generate risk warnings contextually~\cite{chen2025clear}, or even in advance~\cite{slate2025iterative}. Future work could develop models that accurately infer potential harms based on context~\cite{slate2025iterative}. For instance, an agent could distinguish that sharing location data with a mapping provider differs fundamentally from sharing it with an ad network, even if the policy language is similar. This requires creating datasets and conducting evaluation for contextual risks, enabling agents to adapt dynamically.

\textbf{Ensuring Legal Fidelity and Grounding.} The adoption of generative interfaces in legal contexts is complicated by the risk of hallucination and factual error~\cite{zhang2025privcaptcha}. Future work could prioritize reliability, through developing robust RAG frameworks that ground AI responses in source documents and acknowledge uncertainty when policies are ambiguous~\cite{ahmad2020policyqa}. We therefore call for interdisciplinary research to validate conversational strategies that simplify concepts without sacrificing legal precision.

\section{Finding 4: Balancing Stakeholder Opinions}\label{sec:multi_stakeholder}

\begin{figure*}[!htbp]
    \centering
    \includegraphics[width=\textwidth]{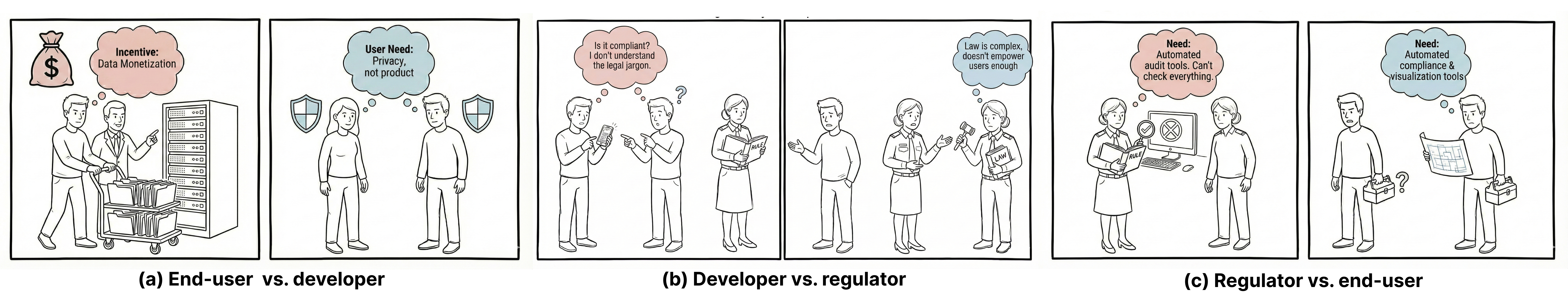}
    \caption{The illustration of the tensions between regulators, users and developers. (a) The \textit{incentive misalignment} for \textit{end-user vs. developer}, (b) Tension in \textit{operationalizing compliance} for \textit{developer vs. regulator}, and (c) \textit{Usability and trust} issues for \textit{regulators vs. end-users}. (The pictures were generated using Gemini Nano Banana and edited by the authors.)}
    \label{fig:finding_4}
    \Description{Three cartoon panels illustrating the tensions between different stakeholders in the privacy ecosystem. (a) End-user versus developer, where a user looks confused and guarded while developers push a cart full of servers and money bags, symbolizing the conflict between user privacy and data monetization. (b) Developer versus regulator, where developers holding mobile devices stand opposite a regulator holding a rulebook, depicting the friction of operationalizing compliance. (c) Regulator versus end-user, where a regulator uses a magnifying glass to inspect a user who is holding a complex newspaper, illustrating the gap between regulatory mandates and actual user comprehension or trust.}
\end{figure*}

The evolution of privacy policy visualization should be understood as the outcome of a complex negotiation among three primary stakeholders: \textit{policymakers and regulators}, who mandate legal compliance; \textit{developers and businesses}, who operationalize these requirements within technical and commercial constraints; and \textit{end-users}, who demand transparency and control (see Figure~\ref{fig:finding_4}). Achieving sustainable progress in privacy requires a holistic approach that balances the tensions among these competing interests, rather than optimizing for any single dimension in isolation.

\subsection{\textcolor{color4}{Developers vs. End-Users: Incentive Misalignment}}

A fundamental barrier to effective privacy transparency lies in the conflict of interest between data-collecting entities (developers and businesses) and end-users~\cite{jones2017probing}. Organizations whose business models rely on data monetization face an inherent misalignment of incentives, lacking the motivation to generate comprehensive, highly legible privacy disclosures~\cite{quach2022digital}. This conflict may result in strategic obfuscation, where enhancing policy legibility can be perceived as a strategic disadvantage by potentially reducing the volume or scope of collected data~\cite{tahaei2023embedding}. 

This business model contribute to a persistent power asymmetry between developers and users~\cite{mohammadi2019pattern}. Although the prevailing \textit{notice-and-control}~\cite{solove2013privacy}\cite{nissenbaum2004privacy} model assumes symmetrical rights, users are often pressured to consent to broad data collection in exchange for access to necessary services. Consequently, developers may lack the internal motivation to ensure the accuracy and precision of both policy text and visualizations~\cite{bednar2019engineering}. This disconnect manifests as widespread non-compliance, with research noting that many applications fail to disclose critical practices, such as location tracking, despite regulatory obligations~\cite{zimmeck2019maps}. This upstream commercial conflict fundamentally undermines efforts focused on improving the downstream usability of privacy interfaces~\cite{kelley2009nutrition,kelley2010standardizing,kelley2013privacy}.

\subsection{\textcolor{color4}{Developers vs. Regulators: Operationalizing Compliance}}

Developers serve as the critical bottleneck in translating abstract legal requirements into functional system practices and user-facing disclosures. Regulators often issue mandates (such as GDPR~\cite{eu_gdpr_2016_full} and CCPA~\cite{ccpa_california_2025}) that require clear, granular disclosure, but these laws frequently rely on abstract language that developers find difficult to interpret or implement efficiently. Even with regulatory enforcements in place, developers, whose core function is often focused on data acquisition~\cite{grover2024encoding}, may seek methods to circumvent or minimize compliance efforts. 

A major operational challenge is the perceived burden of privacy compliance, which developers often view as a secondary task divorced from their core responsibilities~\cite{li2022understanding}. This perception is compounded by practical limitations that developers may lack the expertise or resources to author clear policies~\cite{horstmann2024mapping} or verify data practices embedded within third-party libraries~\cite{shipp2020private}. Beyond privacy policy visualization, Balebako et al.~\cite{balebako2014privacy} already found that developers in smaller companies exhibit less positive privacy behaviors. In IoT domains, technical implementations of product teams also deprioritized privacy, and highlighted national security and consumer trust~\cite{he2025privacy}. To mitigate this burden and empower developers, technical tools have been proposed to integrate privacy checks into development workflows to empower developers--such as systems for formal policy analysis~\cite{da2016improving} or integrated development environment (IDE) tools like \textit{Coconut}~\cite{li2018coconut} and \textit{Matcha}~\cite{li2024matcha}. However, while these tools are a vital step, the persistent challenge~\cite{franke2024exploratory} suggests that more efforts are needed for effective privacy implementation.

\subsection{\textcolor{color4}{Regulators vs. End-Users: Usability and Trust}}

The effectiveness of privacy policy visualization ultimately hinges on its capacity to bridge the gap between legal precision and user comprehension. While legislation like the GDPR~\cite{eu_gdpr_2016_full} mandates clear and accessible notices, the technical or legal definitions of concepts (e.g., metadata~\cite{tang2021defining}) are often complex and ambiguous to convey with technology alone, especially towards end-users. Correspondingly, privacy policies are frequently ignored or abandoned by users due to their obscurity and length~\cite{obar2020biggest,reinhardt2021visual}. Although privacy policy visualization were proposed to act as this bridge~\cite{schaub2015design,schaub2017designing}, how to precisely deliver this data practice information is equally important. While regulation is a powerful and necessary driver for widespread visualization adoption~\cite{barth2022understanding}, its success depends entirely on whether the implementation and visualization techniques can successfully translate the legal framework into understandable information that fosters user trust~\cite{rajaretnam2012problem}.  Furthermore, a lack of understanding prevents law from effectively protecting users, especially who often cannot discern whether an application's stated practices align with regulatory requirements.

To address the scalability required to maintain diverse policy notices (e.g., machine-readable formats, short-form texts, icons)~\cite{schaub2015design, cranor2002web, kelley2009nutrition}, technological efforts have focused on automation. However, automated systems, including advanced LLMs may be subject to hallucinations~\cite{latif2025hallucinations}, and producing inaccurate results~\cite{zhang2025privcaptcha}. Ensuring that automated disclosures are both comprehensible and legally precise requires close collaboration between conversational designers and legal experts~\cite{kolagar2023experts,kim2025interpretation}.

Consequently, user-empowered solutions are potentially a new direction to compensate for the insufficiency of regulatory enforcement and technical limitations. Tools like Collusion (Lightbeam)~\cite{Mozilla_Addons_2019_Lightbeam} and the community-driven \textit{Terms of Service; Didn't Read} (ToS;DR)~\cite{tosdr2012}, as well as risk-based assessments~\cite{chen2025clear} and dialogue systems~\cite{freiberger2025you}, may function to empower users with better self-assessment and negotiation leverage.

\subsection{\textcolor{color4}{Future Directions: Integrating Technology, Policy, and Stakeholders' Practices}}

The future of privacy policy visualization relies on mediating tensions within the socio-technical systems involving regulators, developers and users. Leveraging AI capabilities can shift this static compliance to dynamic, integrated ecosystem management.

\textbf{Scalable Regulatory Oversight and Adaptive Compliance.} A barrier to effective privacy protection is the developer's deprioritization of privacy work~\cite{balebako2015impact,zimmeck2019maps}. Future research should move beyond viewing compliance as a static checkpoint. Instead, we envision AI-powered co-pilots embedded within IDEs to assist developers by automatically interpreting third-party library behaviors~\cite{li2018coconut} and generating legally grounded notices in real time~\cite{chang2024systematic}. For regulators, the integration of LLMs offers a solution to the scalability problem in auditing. Unlike traditional classifiers that require extensive retaining when laws change, LLM-driven auditors~\cite{guo2025repoaudit} may adapt to evolving legal frameworks (e.g., from GDPR~\cite{regulation2016regulation} to the AI Act~\cite{madiega2021artificial}) via updated prompts or knowledge retrieval, eliminating the need for architectural reconstruction. This enables regulators to monitor developer behaviors as an ecosystem scale without lagging behind the pace of legislative updates. 

\textbf{Facilitating Direct Regulator-User Dialogue.} Technology should serve as a bidirectional bridge between the lawmaker and the end-user. While existing work focuses on analyzing policies~\cite{guo2020poli,contissa2018claudette}, future systems could leverage LLMs to perform real-time, context-aware translation of complex legal texts into user-accessible language~\cite{ahmad2020policyqa,ravichander2019question}. More critically, research should explore embedded feedback channels that allow users to report directly to policymaker. By integrating ``regulatory reporting'' mechanisms into app stores or privacy interfaces, users could flag confusing terms or suspicious data practices directly to supervisors. This form of crowdsourced surveillance would provide regulators with granular, real-time insights into systemic violations, fostering a collaborative governance model~\cite{WP29Transparency}.

\textbf{Rebalancing Power Dynamics Between Users and Developers.} Historically, users are forced to rely on developers to visualize privacy risks effectively~\cite{nissenbaum2004privacy,nissenbaum2011contextual}. LLMs may reshape this dynamic by enabling client-side interpretation of privacy policies~\cite{freiberger2025you}. Future research could focus on empowering users with personal privacy agents that autonomously ingest and summarize policies~\cite{harkous2016pribots}, reducing the user's dependency on the service provider's design choices. This means users would no longer need to wait for developers to implement visualizations~\cite{li2022understanding}, but can employ their own tools according to their specific literacy levels and needs, thereby translating the control to the data subject, potentially mitigating dark patterns~\cite{mathur2019dark} or vague disclosures. Furthermore, to bridge the gap between stated policies and actual data practices~\cite{balebako2014privacy}, research could advance towards dynamic auditing and verifiable transparency. For users, this transformation from trust to verification is especially crucial in opaque IoT ecosystems~\cite{guo2013internet}. Such frameworks would not only alleviate consent fatigue~\cite{schaub2015design} but also standardize compliance validation.

\section{Summarization of Open Challenges}\label{sec:discussion}

Based on our synthesis of privacy policy visualization's evolution, we identify four critical and interconnected directions for future research. These directions aim to address the persistent gaps between the potential of privacy policy visualizations and their practical impact, pushing the field towards solutions that are engaging~\cite{zhang2025privcaptcha}, contextually appropriate~\cite{pan2024new,windl2022automating}, inclusive~\cite{feng2024understanding}, and integrated into the real-world software development system~\cite{li2022understanding}.

\begin{table*}[h]
\centering
\caption{A mapping of open challenges, our findings and guidelines.}
\label{tab:guidelines_mapping}
\begin{tabular}{p{0.20\linewidth}p{0.23\linewidth}p{0.52\linewidth}}
\toprule
\textbf{Open Challenge} & \textbf{Findings} & \textbf{Guidelines} \\
\midrule
\textbf{C1: Improving Engagement and Practicality} & \textcolor{color1}{\textbf{F1: Trade-off between Information Load and Decision Efficacy}} & $\bullet$ Use AI for dynamic, personalized policy intervention. \newline
$\bullet$ Consider risk-based notifications. \newline
$\bullet$ Deploy nudging to maximize decision efficacy. \\ \hline

\textbf{C2: Tailoring Visualization to Devices in Different Contexts}  & \textcolor{color3}{\textbf{F3: Tension between Generality and Specificity}} & $\bullet$ Innovate native modalities for emerging platforms. \newline
$\bullet$ Move from adaptation to contextual, multimodal designs. \newline
$\bullet$ Explore haptic, auditory, or ambient visualization cues.
 \\ \hline

\textbf{C3: Adaptive Privacy Policy for Different Demographics}  & \textcolor{color3}{\textbf{F3: Tension between Generality and Specificity}} & $\bullet$ Adapt modality, density, and language complexity. \newline
$\bullet$ Tailor designs based on user literacy and profile. \newline
$\bullet$ Focus on underserved and vulnerable user groups. \newline
$\bullet$ Use LLM-powered systems for content adaptation.
 \\ \hline

\textbf{C4: Integrating Privacy Policy Visualization to Production Environments}  & \textcolor{color2}{\textbf{F2: Co-evolutionary Dynamic of Design and Automation}} \newline \textcolor{color4}{\textbf{F4: Balancing Stakeholder Opinions}} & $\bullet$ Forge developer-technology-regulation links for compliance. \newline
$\bullet$ Create AI-powered co-pilots within IDEs. \newline
$\bullet$ Build socio-legal-technical systems with legal rigor. \newline
$\bullet$ Advance grounded auditing for verifiable compliance. \newline
$\bullet$ Design tools for all stakeholders (regulators, developers).
 \\
\bottomrule
\end{tabular}
\end{table*}

\subsection{Challenge 1: Improving Engagement and Practicality}

A persistent challenge is that privacy policies are frequently ignored due to habituation and a perceived lack of control~\cite{jensen2004privacy, mcdonald2008cost, obar2020biggest}. While novel visualizations exist, effective policy presentation faces an inherent trade-off: \textbf{first, highly engaging formats (e.g., serious games) face practical scalability limits for everyday use~\cite{stellmacher2022escaping, jung2023art}, and second, standardized, static formats (e.g., nutrition labels) hardly overcome user apathy or the sense of ``informed helplessness''~\cite{steinfeld2016agree, kelley2009nutrition, reinhardt2021visual}.} This reflects a broader tension between how much information can be shown and how effectively users can turn that information into decisions.

To overcome these challenges, future research could focus on \textbf{(1) integrated interactive mechanisms}, where policy comprehension is embedded directly into user workflows in an interactive manner to ensure attention, and spur reflective thinking~\cite{terpstra2021think}. The PrivCAP system, which utilizes a mandatory CAPTCHA-like task for policy learning, exemplifies this approach of ``subtle intervention''~\cite{zhang2025privcaptcha}. Future work should expand on this by embedding policy engagement into essential interactions—bridging the gap between the high engagement of games and the utility of standard notices. \textbf{(2) context-aware dynamic visualization}, where the focus shifts from one-off static disclosures to continuous, risk-based dialogue. Research could explore lightweight gamification or dynamic privacy nudges~\cite{wang2013privacy} triggered at critical decision-making moments. With the convergence of policy text and the just-in-time notifications and controls ~\cite{pan2024new}, these interventions can maximize decision efficacy while minimizing user burden.

\subsection{Challenge 2: Tailoring Visualization to Devices in Different Contexts}

While initial privacy policy visualization focused primarily on websites and browsers~\cite{kelley2009nutrition,reinhardt2021visual}, the field has seen the shift toward mobile devices~\cite{zhang2025privcaptcha}, IoT devices~\cite{thakkar2022would} and immersive environments~\cite{lim2022mine}. Transitioning to these platforms introduces two challenges: \textbf{first, the difficulty of effectively communicating policy terms on screenless devices where data collection is passive and inconspicuous~\cite{curran2018your,fruchter2018consumer}, and second, the complexity of negotiating and presenting privacy policy information in specific (e.g., restricted, immersive or shared) environments~\cite{kojic2019impact,naeini2017privacy,zheng2018user}.} In both cases, generic, one-size-fits-all labels struggle to account for device capabilities and social context.

To overcome these challenges, future research could focus on (1) \textbf{ambient and multimodal policy delivery}, addressing the ineffectiveness of traditional visualization on screenless IoT devices~\cite{calo2011against}. Research could move toward ambient information systems~\cite{pousman2006taxonomy} that leverage voice, haptics, or subtle visual cues (e.g., light indicators~\cite{thakkar2022would}) to convey data practices. A critical objective is balancing the efficacy of these policy notifications against user annoyance and alert fatigue~\cite{guo2013internet}. (2) \textbf{multi-party privacy policy content negotiation}, acknowledging that IoT privacy is inherently spatial and collective~\cite{naeini2017privacy, zheng2018user}. Since current labels~\cite{jin2022exploring} often fail to account for multiple users, future systems must determine whose policy preferences take precedence in shared spaces involving household members and guests. Research is needed to visualize these negotiations and transparently communicate how policy terms apply to different occupants~\cite{zhou2024bring}. (3) \textbf{immersive and native policy integration}, where VR platforms require moving beyond placing 2D policy text in 3D space, which suffers from poor readability~\cite{kojic2019impact}. Building on explorations of role-play-based tutorials~\cite{lim2022mine}, future work could design privacy communication native to the immersive environment, for example, by addressing how to represent invisible backend collections and policy content spatially, ensuring that these novel presentations provide genuine clarity rather than a misleading sense of control~\cite{brandimarte2013misplaced}.

\subsection{Challenge 3: Adaptive and Customized Visualization for Different Demographics}

The prevalent ``one-size-fits-all'' approach to privacy policy design inevitably neglects users with diverse literacy levels ~\cite{bhave2020privacy}, decision-making personas~\cite{dupree2016privacy}, and sensory capabilities~\cite{feng2024understanding}. However, shifting to customized design presents two challenges: \textbf{first, the challenge of accurately inferring user context and capability without intrusive profiling, and second, the difficulty of dynamically transcoding legal text into simplified modalities without compromising consistency.}

To overcome these challenges, future research could focus on (1) \textbf{granular user modeling}, where systems could model users' needs and constraints. For example, rather than simply simplifying text for children, research could explore how visualization semantics can map to a child's development stage. Similarly, for visually impaired users, the focus could shift from simple screen-reader compatibility to semantic audio summarization~\cite{feng2024understanding}. (2) \textbf{generative and malleable interfaces}, where recent advancements in LLMs offer the potential to dynamically restructure content based on the user model~\cite{min2025malleable,shaikh2025creating}. A system could generate a high-level iconographic overview for a novice user~\cite{habib2021toggles} while simultaneously synthesizing a detailed data-flow diagram for a technical expert~\cite{reinhardt2021visual}, both derived from the same underlying policy document. Research is needed to validate the accuracy and consistency of these real-time generations. (3) \textbf{conversational scaffolding and dynamic calibration}, where future interfaces could dynamically tune complexity of explanations and the agent's persona based on user feedback~\cite{ait2023power}. This could adapt to the users with diverse mental states and technical literacy. However, this direction requires rigorous evaluation of the ``human-likeness'' of such agents to avoid the uncanny valley effect~\cite{seyama2007uncanny}, ensuring the adaptations enhance comprehension rather than inducing discomfort.

\subsection{Challenge 4: Integrating Privacy Policy Visualization to Production Environments}

Despite a wealth of innovative research proposals for privacy policy visualization~\cite{tabassum2018increasing, reinhardt2021visual}, there remains a significant gap between academic prototypes and their deployment in production environments. Bridging this gap requires addressing the software supply chain and application ecosystems, raising two challenges: \textbf{first, the technical and cognitive burden on developers to create privacy policy visualizations ~\cite{balebako2015impact, li2022understanding}, and second, the lack of incentives for data controllers to voluntarily integrate transparency and control mechanisms within commercial ecosystems.}

To overcome these challenges, future research could focus on (1) \textbf{developer-centric automation and control integration}, acknowledging that user control is fundamentally dependent on developer implementation~\cite{pan2024new,reinhardt2021visual}. Since developers often struggle to engineer the necessary API endpoints for user control~\cite{zimmeck2019maps}, future tools, such as AI-powered IDE ``co-pilots''~\cite{li2024matcha, li2018coconut}, should not only generate legally-grounded notice text but also automatically scaffold the code interfaces required for users to exercise that control. This embeds privacy directly into the development lifecycle. (2) \textbf{client-side agency and negotiation}, which empowers users when providers fail to adopt transparent designs. Research is shifting toward equipping users with independent tools, such as LLM-powered browser extensions and personal privacy agents~\cite{freiberger2025you,chen2025clear}, that can parse policies and summarize terms without the provider's cooperation. This approach effectively uses AI to independently extract transparency from the client side. (3) \textbf{reliability in AI-mediated transparency}, where the reliance on general-purpose models for policy analysis introduces new risks. As research moves toward client-side automation, it must address the potential for hallucinations~\cite{zhang2025siren} and legal misinterpretation~\cite{latif2025hallucinations}. Future work should move beyond simple manual prompts by developing specialized, legally grounded agents that can provide verifiable, accurate, and consistent AI-mediated explanations.

\section{Conclusion}

This paper presented a scoping review of 65 top-tier publications to chart the temporal evolution of privacy policy visualization, addressing the persistent gap between legal compliance and user comprehension. By analyzing the literature through a four-stage design lifecycle framework including context, requirements, design, and evaluation, we identified four dynamic patterns that drive the field's trajectory: (1) the shift from augmenting information presentation to managing information load for improved decision efficacy, (2) the co-evolutionary dynamic where design drive automation, and conversely, advancements in NLP and LLMs unlock new interaction paradigms, (3) the tension between generalized standards and context-specific adaptations in emerging environments, and (4) the requirement to balance conflicting stakeholder incentives for deployment. This work synthesizes these historical insights into actionable guidelines, providing the HCI community with four directions to advance from static visualization to intelligent, user-empowered privacy policy visualization agents.

\begin{acks}
    We thank Prof. Florian Schaub, Yuanyuan Wu and all the reviewers for their help on this paper.
\end{acks}

\bibliographystyle{ACM-Reference-Format}
\bibliography{sample-base}

\appendix 

\section{Search Process, Selection Criteria and Coding Process}\label{app:search_selection}

\begin{figure}[!htbp]
    \includegraphics[width=0.5\textwidth]{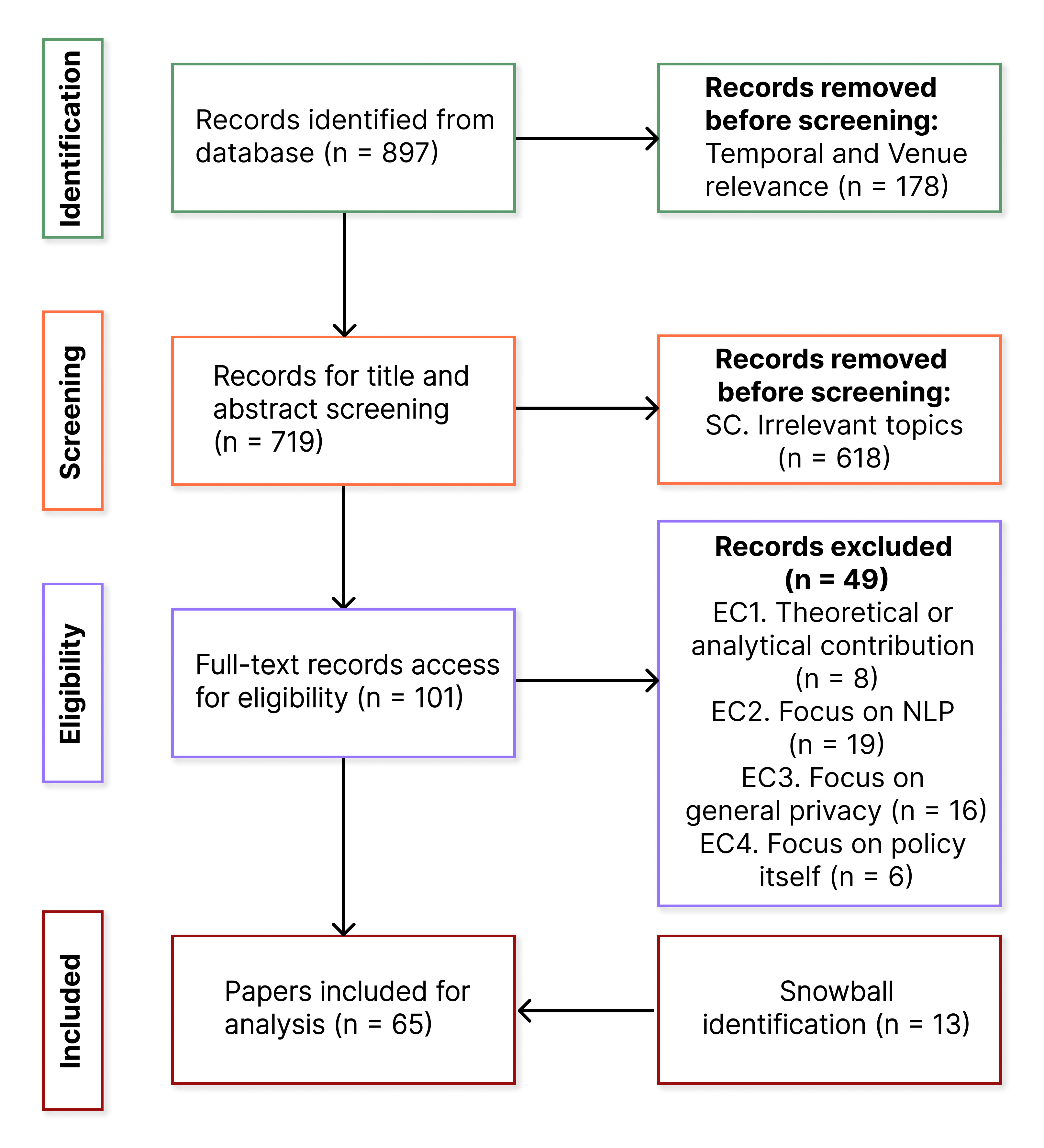}
    \caption{The PRISMA-ScR style literature review process, containing paper selection, screening, eligibility stages.}
    \label{fig:prisma}
    \Description{A PRISMA flow diagram detailing the systematic literature selection process for the review. The initial Identification phase began with 897 records identified from databases, from which 178 were removed before screening. The Screening phase then processed the remaining 719 records for title and abstract review, resulting in the removal of 618 records deemed irrelevant to the topic. The Eligibility phase assessed the full text of 101 records. A total of 49 records were excluded during this stage based on four exclusion criteria: 8 were excluded for being primarily Theoretical, 19 for focusing on Natural Language Processing, 16 for addressing General Privacy topics, and 6 for focusing on Policy content instead of Visualization. Finally, the Included phase incorporated the remaining eligible papers, which were supplemented by 13 papers identified via snowballing, yielding a total of 65 papers included for the final analysis.}
\end{figure}

Our literature review followed the PRISMA-ScR guideline \cite{tricco2018prisma}, and the whole process is outlined in Figure~\ref{fig:prisma}. 

\textbf{Literature Search Details.} Our query string contained the uppercase and lowercase version of the search keywords, applied to the title and content. The terms privacy policy and privacy, policy separated, were iteratively optimized to acquire more relevant papers, as not all papers related to privacy policies could easily be searched with a keyword ``privacy policy''. We have attempted to add ``privacy notice'', ``terms of service'' or other words but did not identify additional eligible papers, only with more irrelevant papers. Therefore, we did not include these keywords in the final search. We selected top-tier HCI conferences including CHI, UIST, CSCW, Ubicomp, IUI, DIS, journals including IJHCI, IJHCS, TOCHI, security conferences including SOUPS, Usenix security and ACM CCS. We scoped our search in top-tier HCI and security \& privacy conferences. 

\textbf{Exclusion Criteria Definitions.} We provided the following clarification on the exclusion criteria. 

\textit{Screening Criteria (Applied to Title and Abstract)}

\textbf{SC-Irrelevant topic:} We excluded papers whose titles or abstracts clearly indicated a primary focus outside the scope of privacy policy visualization, such as policy documents, those with bystander privacy, social negotiations, etc. 

\textit{Eligibility Criteria (Applied to Full Texts)}

\textbf{EC1-Theoretical or analytical contribution.} We excluded those contributing purely on theoretical or analytical work, such as proposing frameworks for the interaction between human and policy documents, just analyzing users' reading cost, etc.

\textbf{EC2-Focus on NLP.} We excluded those focusing on NLP contributions, such as automatic generating privacy policies, automatic analyzing privacy policies. Notably, although we excluded them in our corpus, we do relate these works in the analysis to showcase our findings, such as the spiral law between automation and privacy policy design.

\textbf{EC3. Focus on general privacy.} We excluded those only focusing on general privacy, including bystander privacy, prediction of privacy attitudes, privacy negotiation, etc. Those work did not focus on visualization nor the privacy policy, although they may contain words such as privacy policy in their article.

\textbf{EC4. Focus on policy itself.} We excluded those paper that focused on the privacy policy itself, such as analyzing privacy policies' readability through machine-defined criteria, or through large-scale analysis, as these papers are not related to privacy policy visualization, and did not contribute to our temporal analysis.

\section{Detailed Codebook and Analytical Dimensions}\label{app:codebook}

To foster reproducibility and clarify the boundaries of our analysis, this appendix details the codebook developed into our hybrid deductive-inductive process. The coding scheme is organized into four distinct analytical dimensions, mapping the logical progression from problem identification to solution validation.

\subsection{Dimension I: Context of Use}

This phase focuses on identifying the environmental, regulatory, and user-specific constraints that defines the problem space for privacy policy visualization.

\begin{itemize}

    \item \textbf{Target Audience:} Specifies who the visualization is designed for. Inspired by Mcdonald et al.~\cite{mcdonald2022privacy}, the target audience could involve but is not limited to: 
    \begin{itemize}
        \item \textit{General end-users:} Laypeople with average digital literacy.
        \item \textit{Vulnerable populations:} Children, elderly, or visually impaired users.
        \item \textit{Developers/data controllers:} Tech-savvy users responsible for implementing privacy.
    \end{itemize}

    \item \textbf{Device \& Platform Context:} The specific hardware or software environment targeted, which imposes constraints on the design (e.g., screen size). Inspired by Herriger et al.~\cite{herriger2025context}, the device and platform context could involve but is not limited to: 
    \begin{itemize}
        \item \textit{Web/Desktop:} Traditional browser-based environments.
        \item \textit{Mobile:} Smartphone app ecosystems (iOS/Android).
        \item \textit{IoT/Smart Home:} Connected devices, often screenless (e.g., smart speakers).
        \item \textit{Immersive VR/AR:} 3D or spatial computing environments.
    \end{itemize}
\end{itemize}

\subsection{Dimension II: User Requirements}

This phase analyzes the technical enablers and contribution types necessary to fulfill the identified needs, focusing on the co-evolution of design and automation.

\begin{itemize}
    \item \textbf{Problem Definition (User Challenge):} Operationalizes the specific ``formative findings'' or pain points the paper aims to solve. Synthesizing commonly identified issues from prior work~\cite{obar2020biggest,reinhardt2021visual}, the challenges could involve but are not limited to: 
    \begin{itemize}
        \item \textit{Information asymmetry:} Challenges related to the opacity of data practices.
        \item \textit{Cognitive overload:} Issues regarding the length, complexity or readability of policy text.
        \item \textit{Reluctance:} user apathy or ``privacy fatigue'' leading to ignored policies.
        \item \textit{Contextual decoupling:} The disconnect between static and dynamic data usage.
    \end{itemize}
    
    \item \textbf{Contribution Type:} Classifiers the primary research output of the paper. Informed by Wobbrock et al.~\cite{wobbrock2012seven,wobbrock2016research}, the contribution type includes:
    \begin{itemize}
        \item \textit{Empirical Study:} Research focused on presenting new findings derived from systematically collected data.
        \item \textit{Artifacts:} Interventions within HCI, such as new systems, architectures, tools, techniques, or designs. These are valued for revealing new opportunities, enabling novel outcomes, or facilitating future designs and explorations.
        \item \textit{Methodology:} Contributions evaluated primarily based on the utility of a new or significantly improved research method.
        \item \textit{Theoretical:} The development of new or improved concepts, definitions, models, principles, or frameworks.
        \item \textit{Benchmark/Dataset:} The provision of a much-needed corpus for the research community, enabling the testing of future innovations.
        \item \textit{Survey:} Work that reviews and synthesizes existing research in a field to expose key trends, themes and gaps. 
        \item \textit{Opinion:} Papers that aim to influence the reader through persuasive argumentation on a specific topic. 
    \end{itemize}
\end{itemize}

\subsection{Dimension III: Design Solutions}

This phase captures the specific visualization artifacts, content and interaction patterns proposed.

\begin{itemize}
    \item \textbf{Visualization Metaphor:} The core design concept used to translate legal text into visual form. This includes but is not limited to: 
    % {\color{blue}terminology, icons, stories, data-flow diagrams, and conversations, haev been used in recent study of privacy labelling \cite{barth2022understanding}}
    \begin{itemize}
        \item \textit{Standardized Label/Table:} Nutrition label-style grids or structured tables~\cite{kelley2009designing,kelley2009nutrition}.
        \item \textit{Iconography:} Visual symbols representing specific data practices~\cite{harkous2018polisis,gerl2018extending}.
        \item \textit{Narrative/Comic:} Story-based or illustrated explanations~\cite{tabassum2018increasing}.
        \item \textit{Data Flow/Graph:} Network visualizations of data movement~\cite{guo2020poli}.
        \item \textit{Conversational Interface:} Chatbots or voice-based assistants~\cite{chen2025clear,freiberger2025you}.
    \end{itemize}

    \item \textbf{Information Content:} The specific data practices presented in the visualization. Inspired by prior work~\cite{reinhardt2021visual,zhang2025privcaptcha}, the information content dimension includes but is not limited to:
    \begin{itemize}
        \item \textit{Data Types:} What data is collected (e.g., location, contacts).
        \item \textit{Purpose:} Why data is collected (e.g., advertising, functionality).
        \item \textit{Sharing/Third-parties:} Who the data is shared with.
        \item \textit{User Rights/Control:} Options for opt-out, deletion or access.
    \end{itemize}

    \item \textbf{Automation \& Technology Level:} Codes the specific technical methods used to generate the visualization content. The automation type includes but is not limited to:
    \begin{itemize}
        \item \textit{Manual/Static:} Human-authored summaries or labels.
        \item \textit{Rule-based/Protocol:} Use of structured standards like P3P.
        \item \textit{NLP/ML Extraction:} Use of NLP or ML classifiers to extract practices.
        \item \textit{LLM/Generative AI:} Use of LLMs for dynamic synthesis or Q\&A.
    \end{itemize}
\end{itemize}

\subsection{Dimension IV: Evaluation}

This phase examines how the design solutions were validated and what metrics were used to measure success.

\begin{itemize}
    \item \textbf{Evaluation Methodology:} The study design used to validate the visualization. Inspired by established practices~\cite{dutt1994evaluating,barkhuus2007mice} and refining specific within our visualization scope, this includes but is not limited to:
    \begin{itemize}
        \item \textit{User Study (Lab/Online):} Controlled experiments or surveys.
        \item \textit{Field Study/Deployment:} In-the-wild testing.
        \item \textit{Expert Review:} Heuristic evaluation or expert interviews. 
        \item \textit{Automated Analysis:} Benchmarking against datasets (common in NLP papers).
    \end{itemize}

    % \item \textbf{Success Metrics (Measurements):} The specific dependent variables to assess effectiveness. According to the past guidelines on evaluating privacy choice mechanisms~\cite{cranor2023metrics,habib2022privacychoice}, this includes but is not limited to the following:
    % \begin{itemize}
    %     \item \textit{Comprehension/Accuracy:} Can users correctly answer questions about the policy?
    %     \item \textit{Efficiency/Time:} How long does it take to find information?
    %     \item \textit{Behavior/Decision:} Does it change installation rates or permission granting?
    % \end{itemize}
\end{itemize}

Besides each stage, we also coded some overall dimensions, aiming to track temporal and disciplinary trends:

\begin{itemize}
    \item \textit{Year:} The publication year of the corresponding paper.
    \item \textit{Venue:} The publication conference/journal venue (e.g., CHI, SOUPS, USENIX Security).
\end{itemize}

\section{Paper List}

\onecolumn
\begin{scriptsize}
\setlength{\LTcapwidth}{\textwidth}
\begin{longtable}{
    c  % Paper
    c  % Year
    p{2.3cm}  % Venue
    c  % Contribution
    c  % Challenge
    c  % Audience
    p{1.3cm}  % Regions
    c  % Platform
    c  % Format
    c  % Technique
    c  % Content
    c  % Type
}

\caption{The paper list. \textbf{For contribution}, P denotes prototype, E denotes empirical, D denotes design guideline, T denotes theoretical framework, S denotes system. For challenge, P denotes poor readability and comprehension, L denotes low user engagement, U denotes uninformed consent and powerlessness, I denotes ineffective presentation, T denotes lack of transparency. \textbf{For audience}, G denotes general, D denotes developers, L denotes law makers or regulators, R denotes researchers. \textbf{For regions}, NA denotes not specified. \textbf{For platform}, W denotes website, V denotes virtual reality/augmented reality, M denotes mobile, I denotes smart home IoT devices, NA denotes not specified. \textbf{For format}, T denotes tabular-based visualization, G denotes graph-based visualization, D denotes dialogue-based visualization, I denotes interactive visualization,
 C denotes CPP, N denotes notices or labels, V denotes immersive visualization in virtual, augmented environment, or gamified visualization, P denotes layered privacy policy or original privacy policy, NA denotes not specified. \textbf{For technique (i.e., technique involved in the paper)}, NA denotes not specified. \textbf{For content (i.e., visualized content)}, C denotes data collection, D denotes data disclosure, P denotes data processing, S denotes data sharing and usage, O denotes opt-out. \textbf{For type (i.e., experiment type)}, U denotes user study, No denotes no evaluation, NA denotes not specified, T denotes technical evaluation, E denotes empirical measurement.}
\\

\toprule
\textbf{Paper} & \textbf{Year} & \textbf{Venue} & \textbf{Contribution} & \textbf{Challenge} &
\textbf{Audience} & \textbf{Regions} & \textbf{Platform} & \textbf{Format} &
\textbf{Technique} & \textbf{Content} & \textbf{Type} \\
\midrule
\endfirsthead

\toprule
\textbf{Paper} & \textbf{Year} & \textbf{Venue} & \textbf{Contribution} & \textbf{Challenge} &
\textbf{Audience} & \textbf{Regions} & \textbf{Platform} & \textbf{Format} &
\textbf{Technique} & \textbf{Content} & \textbf{Type} \\
\midrule
\endhead

\bottomrule
\endfoot

            ~\cite{reinhardt2021visual}   & 2021 & CHI & P,E,D & L,P & G & Germany & W & T,I & NA & C,D,P & U \\
            ~\cite{nouwens2020dark}       & 2020 & CHI & E & U,I,T & G & Europe & W & N & NA & C,P & U \\
            ~\cite{jung2023art}           & 2023 & CHI & E & P,L,I,T & G & NA & V & V & NA & C & No \\
            ~\cite{griggio2022caught}     & 2022 & CHI & E & P,U & G & MX, ES, ZA, UK & M & P & NA & C & U \\
            ~\cite{shiri2024motivating}   & 2024 & DIS & P,E,T & L & G & Canada & M & I & NA & C,S,O & U \\
            ~\cite{soumelidou2020effects} & 2020 & IJHCI & P,E & P,I,L & G & NA & W & P,G & NA & C,S,O & U \\
            ~\cite{rudolph2018users}      & 2018 & HCII & E,T & L & G & NA & W & P & NA & S,O & U \\
            ~\cite{zhang2025privcaptcha}  & 2025 & CHI & P,E & P,L & G & China & M & I,T & LLM & C,D,P,S & U,T \\
            ~\cite{jensen2004privacy}     & 2004 & CHI & E & I & G & NA & W & P & NA & C,P,S,O,D & U \\
            ~\cite{chen2023ask}           & 2023 & SOUPS & E & U,I & G & US & I & T,N & NA & C,P,D & U \\
            ~\cite{lie2022automating}     & 2021 & TOCHI & S,T & T,I & D & NA & NA & NA & NA & C,S & E \\
            ~\cite{harkous2018polisis}    & 2018 & USENIX & S,T & P,I & G & NA & W & T & NLP & C,S,O & T \\
            ~\cite{guo2020poli}           & 2020 & WPES & P,E & P,L & G & US & W & G & NA & C,P,S & U \\
            ~\cite{ghazinour2016usability}& 2016 & DASC & E & P,L,I & G & NA & W & T & NA & C,O & U \\
            ~\cite{stellmacher2022escaping}&2022 & CHI PLAY & P,E & P,L,U & G & Germany & M & V & NA & C,P,S,O & U \\
            ~\cite{gerl2018extending}     & 2018 & BSC Learning & P,E & I,U & G & Europe & W & P & NA & P,S,O & U \\
            ~\cite{volk2022pricheck}      & 2022 & CHI & E,S & U,I & G & NA & W,I & N,T & NA & C,S & U \\
            ~\cite{brunotte2022my}        & 2022 & ICSSP & P,E & P,T & G & NA & W & P,G & NA & C,S,O & U \\
            ~\cite{kitkowska2023designing}& 2023 & Computers \& Security & E & U,P,I,L,T & G & NA & W & G,P & NA & D,C,S & U \\
            ~\cite{zaeem2018privacycheck} & 2018 & TOIT & P & P,L & G & NA & W & I,P & NLP & D,S,O & T \\
            ~\cite{atashpanjeh2025mental} & 2024 & IJHCI & P,E & P,I,L & G & US,CA & W & G,V & NA & D,S & U \\
            ~\cite{levy2005improving}     & 2005 & WWW & P,E & P,L & G & NA & W & P & XML & C,D,O & U \\
            ~\cite{ortloff2020implementation} & 2020 & DIS & P,E & P,L,I & G & Germany & W & C,N & NA & C,P & U \\
            ~\cite{tsolakidou2023exploring} & 2023 & PCI & E & P,L & G & NA & W & P,G,N & NA & C,S & U \\
            ~\cite{jones2017probing}     & 2017 & EVA & P,E & P,L,U,T & G & UK & W & G & NA & C,D,P,S & U \\
            ~\cite{romanosky2006privacy} & 2006 & PLoP & T & P,T,I & D & US,Europe & W & NA & NA & C,D & U \\
            ~\cite{zhang2024exploring}   & 2024 & SOUPS & T,E,U,I & G & N,T & US & M & N,T & NA & C,P,S & U \\
            ~\cite{srinath2021privaseer} & 2021 & ICWE & S & P,L,T & L,R & US,EU & W & I,D & NLP & C,D,S,O & T \\
            ~\cite{pierce2018interface}  & 2018 & DIS & P,E & P,L,U,T & G & US & W,M & G,V & NA & C,P,S & U \\
            ~\cite{kelley2008user}       & 2008 & AISec & T & P,T & R,D & NA & M & T,N & NA & D,S & T \\
            ~\cite{schaub2017designing}  & 2017 & IEEE Internet Computing & P,E,D & U,L,I & G & NA & NA & N & NA & C,S,O & No \\
            ~\cite{kelley2009nutrition}  & 2009 & SOUPS & E & P,L,U & N,T & NA & NA & NA & NA & C,S & U \\
            ~\cite{xiao2025transparency} & 2025 & IJHCI & E,T & P,T,L,U & G & NA & W,M & P & NA & C,S,O & U \\
            ~\cite{zeng2022privacy}      & 2020 & Journal of Business Ethics & E,T & P,L,T & G & China & NA & P,G & NA & C,D,S & U \\
            ~\cite{chen2025clear}        & 2025 & IUI & S & P,L,U,I & G & NA & W & I,D,C & LLM & C,P,O & U \\
            ~\cite{ghazinour2009model}   & 2009 & International Computer Software and Applications Conference & P & I,P,L & G,L & NA & NA & G & NA & C,P & No \\
            ~\cite{reeder2008user}       & 2008 & WPES & P,E & P,I & G & NA & W & T,N & NA & C,S & U \\
            ~\cite{becker2014effect}     & 2014 & HICSS & P,E & T,P & G & NA & W & P & NLP & C,S & U \\
            ~\cite{lipford2010visual}    & 2010 & CHI & E & P,I & G & NA & W & N,T & NA & S,O & U \\
            ~\cite{reeder2008expandable} & 2008 & CHI & S & T,P,U,I & G & NA & W & T & NA & C,S,D & U \\
            ~\cite{tabassum2018increasing} & 2018 & CHI & P & P,L,U & G & NA & W & G & NA & C,S,O & U \\
            ~\cite{kelley2010standardizing} & 2010 & CHI & P,E & P,L,I & G & NA & W & T,N & NA & C,P,S & U \\
            ~\cite{kelley2009designing}  & 2009 & CHI & P & P & G & NA & W & T,N & NA & C,S,O & U \\
            ~\cite{ayyavu2011integrating}& 2011 & CHI & E & U & G & NA & NA & NA & NA & C,D,S & U,T \\
            ~\cite{taber2020beyond}      & 2020 & CHI & P,E & P,L,U,I & G & NA & W & P & NA & C,D,O & U \\
            ~\cite{kay2010textured}      & 2010 & SOUPS & P,E & L,P,U & G & NA & W & P & NA & C,D,O & U \\
            ~\cite{bannihatti2020finding}& 2020 & WWW & S & I,T,L,U & D & NA & NA & NA & NLP & C,O,S & T \\
            ~\cite{waddell2016make}      & 2016 & CHI & P,E & L,P,I,U & G & NA & W & P & NA & C,D & U \\
            ~\cite{bolchini2004need}     & 2004 & ICWE & S,E & P,I & G & NA & W & C,N & NA & C,S & U \\
            ~\cite{knijnenburg2016comics}& 2016 & SOUPS & P,D & P,L,U,I & G & NA & W & G & NA & D,S,O & U \\
            ~\cite{freiberger2025you}    & 2025 & CHI & P,E & P,I,T & G & NA & W & I,D & LLM & C,S,O & U \\
            ~\cite{lim2022mine}          & 2022 & CHI & S & P,L,I & G & NA & V & V & NA & C,S & U \\
            ~\cite{shanmugarasa2025privacy} & 2025 & CHI & S & L,T & G & NA & W & C & NLP & C,D,S & T,U \\
            ~\cite{stover2021investigating} & 2021 & CHI & D,P,E & P,L,I & G & NA & M & N & NA & C & U \\
            ~\cite{windl2022privacy}     & 2022 & CHI & D,P,E & P,L,I & G & NA & W & N,P & NA & C,O & U \\
            ~\cite{cranor2006user}       & 2006 & TOCHI & P,D & P,U,I & G & NA & W & N & XML & C,S,O & U \\
            ~\cite{zimmeck2014privee}    & 2014 & USENIX & S,E & P,L,T,U & G & NA & W & NA & NLP & C,D & T \\
            ~\cite{angulo2012towards}    & 2012 & Information Management \& Computer Security & D,P & P,I & G & NA & W & T,G & NA & C,S & U \\
            ~\cite{mohammadi2019pattern} & 2019 & EuroPLoP & D,P & P,L,U,I,T & G & NA & W & T & NLP & C,S,D,O & No \\
            ~\cite{chen2017sweetdroid}   & 2017 & WPES & S & U,L,P & G & NA & W & NA & NLP & C,D,S & T \\
            ~\cite{kelley2013privacy}    & 2013 & CHI & P,E & P,L,I,U,T & G & NA & W & T,N & NA & C,D,O & U \\
            ~\cite{habib2021toggles}     & 2021 & CHI & D,E & P,I & G & NA & W & N & NA & C,S,O & U \\
            ~\cite{feng2021design}       & 2021 & CHI & S,P & P,U,I & G & NA & I & T,N & NA & C,P,S,O & U \\
            ~\cite{thakkar2022would}     & 2022 & CHI & D,E & T,U & G & NA & I & T,N & NA & C,S & U \\
            ~\cite{luger2013consent}     & 2013 & CHI & E & T,U,L,P & G & NA & W & NA & NA & C,S & E \\
            \bottomrule
            
\end{longtable}
\end{scriptsize}
\twocolumn

\end{document}